\documentclass[structabstract]{aa}  
\usepackage{graphicx}
\usepackage[longnamesfirst]{natbib}
\usepackage[normalem]{ulem} 
\usepackage{color}
\usepackage{booktabs} 
\usepackage{subfigure}
\usepackage{aalongtable}
\usepackage{txfonts}
\shortcites{Ammler05a}
\shortcites{An07}
\shortcites{Asiain99}
\shortcites{Asiain99err}
\shortcites{Allende04}
\shortcites{BCAH98}
\shortcites{Barklem02}
\shortcites{Biazzo06}
\shortcites{Biemont91}
\shortcites{Biemont93}
\shortcites{Burrows97}
\shortcites{Carlsson94}
\shortcites{Cayrel85}
\shortcites{Chen00}
\shortcites{Chereul99}
\shortcites{Chupina01}
\shortcites{Cyburt03}
\shortcites{deLaverny03}
\shortcites{Delfosse98}
\shortcites{Edvardsson93}
\shortcites{Fuhrmann97a}
\shortcites{Fuhrmann97b}
\shortcites{Gray03}
\shortcites{Hanbury74}
\shortcites{Hatzes05}
\shortcites{Hog00}
\shortcites{Huensch99}
\shortcites{Huensch04}
\shortcites{Kenworthy01}
\shortcites{King00}
\shortcites{King03}
\shortcites{Koenig02}
\shortcites{Koenig05}
\shortcites{Koenig06}
\shortcites{Neuhauser97}
\shortcites{Paulson03}
\shortcites{Piskunov95}
\shortcites{Ryabchikova97}
\shortcites{Kupka99}
\shortcites{Kupka00}
\shortcites{Lowrance05}
\shortcites{Montes01b}
\shortcites{Montes01a}
\shortcites{Nordstroem04}
\shortcites{Neuhaeuser97}
\shortcites{Randich03}
\shortcites{Richichi98}
\shortcites{Perrin98}
\shortcites{Perryman97}
\shortcites{Perryman98}
\shortcites{Pfeiffer98}
\shortcites{Potter02}
\shortcites{Schuler03}
\shortcites{Soderblom90}
\shortcites{Soderblom93a}
\shortcites{Soderblom93c}
\shortcites{Strassmeier00}
\shortcites{Terndrup00}
\shortcites{Thorburn93}
\shortcites{Walter84}
\shortcites{VandenBerg00}
\shortcites{Yi01}

\defcitealias{Fuhrmann04}{F04}
\defcitealias{Montes01b}{M01}
\defcitealias{King03}{K03}

\def \tablinesep {0.1em}
\def \fhour {\hbox{$.\!^{h}$}}
\def \fdeg {\hbox{$.\!^{\circ}$}}
\def \farcs {\hbox{$.\!"$}}

\begin{document}
   \title{Spectroscopic properties of cool Ursa Major group
   members\thanks{Based on observations with the Coud\'e-\'Echelle
   spectrograph of the Alfred-Jensch-Teleskop at the Th\"uringer
   Landessternwarte Tautenburg.}\thanks{Based on observations collected
   at the Centro Astron\'omico Hispano Alem\'an (CAHA) at Calar Alto,
   operated jointly by the Max-Planck-Institut f\"ur Astronomie and the
   Instituto de Astrof\'isica de Andaluc\'ia (CSIC).}}


   \author{M. Ammler-von Eiff\inst{1}\inst{2}\inst{3}\inst{4} \and E.W. Guenther\inst{5}
          }
   \offprints{M. Ammler-von Eiff}

   \institute{Centro de Astronomia e Astrof\'isica da Universidade de Lisboa, Observat\'orio Astron\'omico de Lisboa, Tapada da Ajuda, 1349-018 Lisboa, Portugal 
         \and Centro de Astrof\'{\i}sica da Universidade do Porto, Rua das Estrelas, 4150-762 Porto, Portugal 
         \and Astophysikalisches Institut und Universit\"ats-Sternwarte Jena, Schillerg\"a{\ss}chen 2-3, 07745 Jena, Germany
         \and Georg-August-Universit\"at G\"ottingen, Institut f\"ur Astrophysik, Friedrich-Hund-Platz 1, 37077 G\"ottingen, Germany, \email{mammler@uni-goettingen.de}
         \and Th\"uringer Landessterwarte Tautenburg, 07778 Tautenburg, Germany
             }

   \date{Received September 15, 1996; accepted March 16, 1997}

 
  \abstract
   {Until now, most members of the Ursa Major (UMa) group of stars have
    been identified by means of kinematic criteria. However, in many
    cases kinematic criteria alone are insufficient to ascertain, whether an individual star is really a member of this group. Since photometric criteria are ineffective in the case of cool dwarf members, one must use spectroscopic criteria. Nevertheless, resulting membership criteria are inconclusive.}
   {We reanalyse spectroscopic properties of cool UMa group dwarfs. In particular, we study the distribution of iron abundance, the strength of the Li\,I absorption at $6708\,\mathrm{\AA}$ and the Li abundance, and the infilling of the core of the H$\alpha$ line.}
   {Twenty-five cool and northern bona-fide members are carefully selected from the literature. Homogeneously measured stellar parameters and iron abundances are given for all Sun-like stars selected, based on spectra of high resolution and high signal-to-noise ratio. In addition, we measure the Li equivalent width and abundance as well as the relative intensity of the H$\alpha$ core and the corresponding chromospheric flux.}
   {The studied stars infer an average Ursa Major group iron abundance of $-0.03\pm0.05\,$dex, which is higher by about $0.06\,$dex than determined elsewhere. The Li abundance derived of Ursa Major group dwarf stars is higher than in the Hyades at effective temperatures cooler than the Sun, but lower than in the younger Pleiades, a result which is independent of the exact value of the effective temperature adopted. The Sun-like and cooler dwarfs also display chromospheric infilling of the H$\alpha$ core. We present spectroscopic criteria that may be used to exclude non-members.}
   {}

   \keywords{Stars: abundances  -- Stars: fundamental parameters
                  }
   \titlerunning{UMa group stellar parameters}

   \maketitle
%
\section{Introduction}
\subsection{The UMa group}


In the 19$^\mathrm{th}$ century, \citet{Proctor1869} and \citet{Huggins1871} realized that five of the A stars in the Big Dipper
constellation move to a common convergence point, and are thus likely to move into the same direction of space. In 1909, \citeauthor{Hertzsprung09} identified
stars at a very large angular distance from the Big Dipper constellation
that were also moving towards the same convergent point, and thus had to be members of the same group of stars. All of these stars form an association with a central concentration, the UMa open cluster in the Big Dipper constellation, located at a distance of about 25\,pc. Since the stars belonging to
the UMa group probably do not form a single open cluster
\citep{Wielen78a}, the term {\em Sirius
supercluster} has instead been suggested \citep{Eggen94}, or {\em UMa association}
\citep{Fuhrmann04}, or {\em UMa group}. Since the nature of
this cluster or association of stars is not yet well established,
we prefer to use the more general term {\em UMa group} in this work. A
number of reviews of the UMa group of stars have
already been published: while \citet{Roman49} summarizes work until the middle of the 20$^\mathrm{th}$ century,
\citet{Eggen92} and \citet{SM93} review the pre-{\em
Hipparcos} era, and \citet{Asiain99} the post-{\em Hipparcos}
research. Good accounts are also given by both \citet{King03} and \citet{Fuhrmann04}.

The UMa group of stars is well-suited to many different kinds of studies because of its relative proximity, which enabled the distances and proper motions of its members to be determined
to high accuracy by the {\em Hipparcos} mission
\citep{ESA97,Perryman97}. The UMa group is young, its age being estimated
to be $\approx300\,$Myrs \citep{vonHoerner57, Giannuzzi79, Duncan81,
SM93}\footnote{The range of estimates varies significantly between 200\,Myrs \citep{Koenig02} and 600\,Myrs \citep{KS05}. The age given by \citet{Koenig02} is based on the analysis of the Sun-like member $\chi^1\,$Ori (HD\,39587) and its low-mass companion while the photometric age given by \citet{KS05} is constrained by the main-sequence turn-off of early-type members.}. The young age of the UMa group is close to the time scale of the dissolution of open clusters
\citep{Wielen71}. Studies of its members thus allow one to analyse aspects of stellar and cluster evolution.

\subsection{Kinematic and photometric membership criteria}

Is the lack of an established membership list the reason for the disagreement in the age of the UMa group? 

The definition of kinematic membership criteria for the UMa group has always been a matter of debate. There are basically two different approaches. The first technique defines a list of canonical members, in this case the UMa nucleus, and derives kinematic criteria based on this sample of stars \citep[e.g.,][]{SM93,Montes01b,King03}. \citet{King03} point out that the UMa group kinematics might be biased by the adoption of the canonical list of members.
A second and unbiased approach looks for clustering in the kinematical space, an approach originating from \citet{Dziewulski16}. Such work strongly benefits from the {\em Hipparcos} mission \citep[e.g.,][]{Asiain99,Asiain99err,Chereul99}. 

In addition to kinematic criteria, photometric data can be used, since moving group members were born from the same material at approximately the same time. A photometric criterion was applied to the UMa group by \citet{King03} and \citet{KS05} who required that all members had to be located on the same isochrones throughout the Hertzsprung-Russell diagram implying an age of about 600\,Myrs. The method is effective for the early-type members of the UMa group since the isochrones are then sensitive to age. At cooler effective temperatures, i.e., in the case of solar and later type stars, luminosities and colours are less sensitive to age and the photometric criterion becomes less useful.

\subsection{Chemical homogeneity}

A similar abundance pattern is expected for members of moving groups since they are supposed to form from the same material. Indeed, observations show that stellar groups differ in terms of iron abundance, e.g., \citet{BF90} measured [Fe/H]$=-0.03\pm0.02\,$dex for the Pleiades, and [Fe/H]$=0.13\pm0.05\,$dex for the Hyades.
For the UMa group, they inferred that [Fe/H]$=-0.085$, although a large spread of $0.20\,$dex seems realistic \citep{SM93,GG02}.


When abundance data are compiled from the literature, inhomogeneity is a major complication, as pointed out by \citet{King03}.
The spread of abundance measurements for the same star can be very large as illustrated by the entry of the UMa group member HD\,39587 in the catalogue of \citet{Cayrel01} (Table~\ref{tab:cayrel01}). While the effective temperatures agree well, the surface gravities and abundances scatter widely. Therefore, an interesting question is how much of the iron abundance scatter detected in the UMa group is intrinsic.

\begin{table*}
\caption{\label{tab:cayrel01}Inhomogeneity of literature data illustrated by the entry for the UMa group member HD\,39587 in the catalogue of \citet{Cayrel01}.}
\small
\begin{tabular}{llll}
\toprule
$T_\mathrm{eff}$\,[K]&${\log}(g\,[\frac{\mathrm{cm}}{\mathrm{s}^2}])$&[Fe/H]&Source\\
\midrule
5929     &                 &-0.05         & \citet{Boesgaard88}\\
5929     &                 &-0.05         & \citet{Boesgaard89}\\
5929     &   4.50          &-0.05         & \citet{BF90}\\
5900     &   4.21          &-0.05         & \citet{Friel92}\\
5953     &   4.46          &-0.03         & \citet{Edvardsson93}\\
5895     &   4.21          &-0.04         & \citet{Gratton96}\\
5950     &   4.46          &+0.11         & \citet{Mallik98}\\
5929$\pm$70 &   4.49  $\pm$  0.10  &-0.02  $\pm$  0.12 & \citet{Castro99}\\
5805$\pm$70 &   4.29 $\pm$   0.10  &-0.18 $\pm$   0.10 & \citet{Chen00}\\

\bottomrule
\end{tabular}
\end{table*}

\subsection{Spectroscopic indicators of youth}
The presence of Li\,I absorption at $\lambda\lambda6707.8$\,{\AA} indicates stellar youth \citep[e.g.,][]{Thorburn93,Soderblom93a,Carlsson94}. Young stars also display an enhanced level of activity, which can be spectroscopically observed as chromospheric infilling of the H$\alpha$ line \citep{Herbig85,Martin97} or the Ca\,II H\&K lines, which has already been detected for the UMa group by \citet{Roman49}.


A homogeneous level of both infilling of Balmer lines and Li absorption has also been detected \citep{Montes01a,Fuhrmann04,KS05}. Interestingly, \citet{SM93} noted that UMa group stars selected on the basis of spectroscopy also exhibit closer agreement with the kinematics of the nucleus of the UMa group. In the same fashion, \citet{King03} point out that the dispersions in the space velocities increase from those of the probable activity-based members to those of the activity-based non-members -- a pattern that is also found for the UMa group sequence in the HR diagram.

\citet{Fuhrmann04} highlights that stars isolated from a complete sample of nearby stars based on spectroscopic features of activity, Li absorption, and enhanced projected rotational velocity have similar space velocities to those of the UMa cluster core. 

\subsection{A new uniform study}
However, all this work is based either on only a small number of stars or on inhomogeneous data. So far, the definition of membership criteria beyond kinematics turns out to be difficult. \citet{King03} realize 

\begin{quotation}
"... that activity is considerably more robust at excluding [UMa group] membership"
\end{quotation}

\noindent 
and

\begin{quotation}
"... photometric membership and abundance-based membership are necessary but far from sufficient conditions to guarantee UMa membership."
\end{quotation}


Our new uniform work, which is based on as many late-type UMa group members as possible, will not only allow one to study the level of homogeneity of abundances, Li absorption, and H$\alpha$ infilling. The new data may also help to refine previous approaches to defining spectroscopic membership criteria. 

While the present work rectifies problems caused by inhomogeneous data or small sample numbers, it may still be affected by any problems related to the sample selection, which has to rely on a list of bona-fide members.

\begin{figure}
  \centering
  \subfigure{\includegraphics[width=8cm]{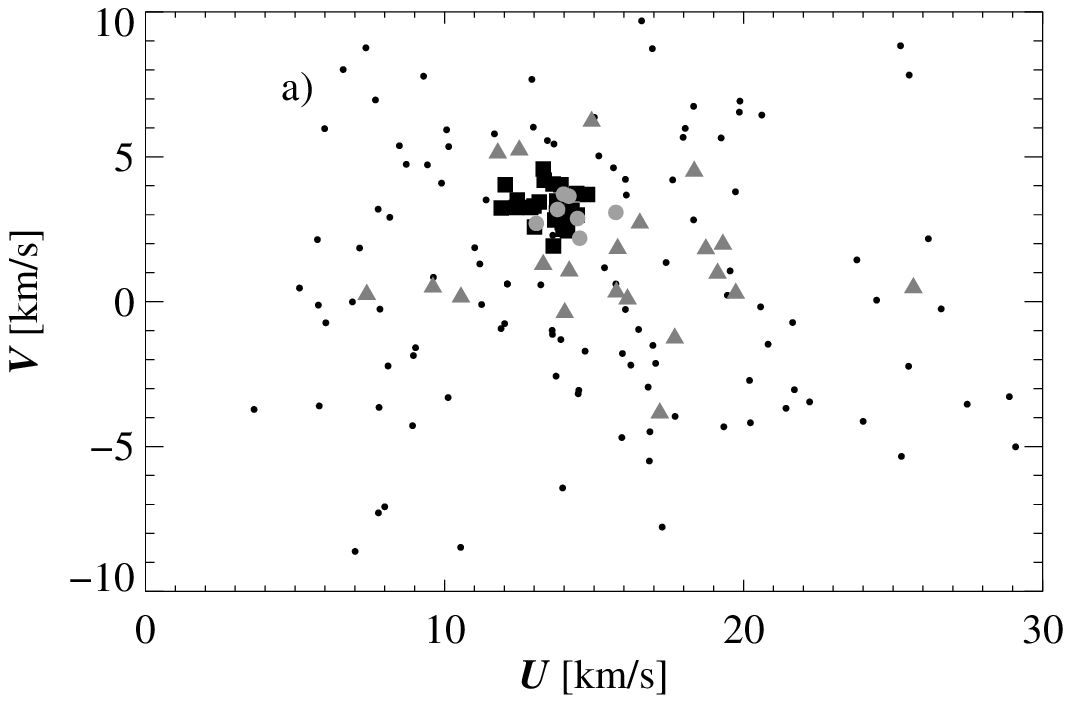}}
  \subfigure{\includegraphics[width=8cm]{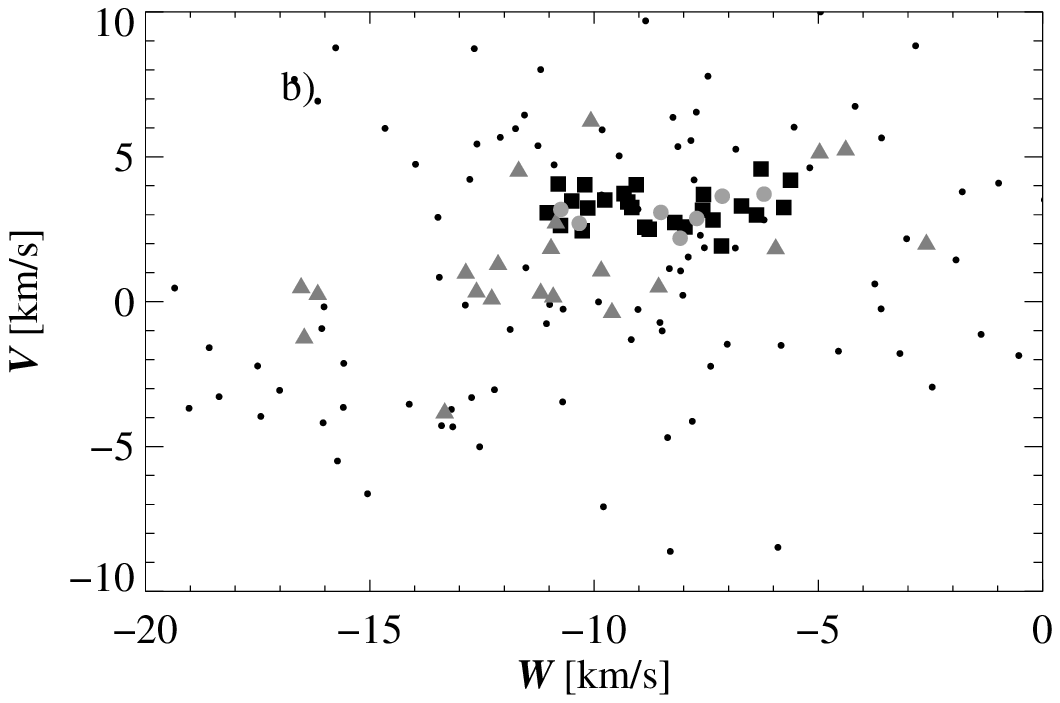}}
  \caption[Space velocities of the kinematic sample]{\label{fig:comp_kin}{\bf Space velocities of the kinematic sample} -- {\bf (a)} The Galactic space velocities U and V of kinematic members and further UMa group candidates are displayed. Triangles indicate the stars satisfying both membership criteria of \citetalias{Montes01b}. Squares denote the stars that agree with the kinematic criterion of \citetalias{King03}. Circles are the seven stars that fulfil the criteria of both \citetalias{Montes01b} and \citetalias{King03}. Dots are additional UMa group candidates taken from literature. {\bf (b)} Same as a), now for $W$ and $V$. Note that the dots do not fulfil both $UV$ and $WV$ constraints simultaneously. If space velocities are given by both \citetalias{Montes01b} and \citetalias{King03}, then they are taken from the former.}
\end{figure}

Table~\ref{tab:uma_kin} compares the kinematic properties of the UMa group derived by different methods. We compare the most recent member lists, i.e, the lists published by \citet{Montes01b}\citepalias{Montes01b} and \citet{King03}\citepalias{King03}, which have 76 stars in common. Twenty-eight of these stars are kinematic members, according to \citetalias{Montes01b}, that fulfil both membership criteria. Sixteen of the 76 stars are kinematic members, according to \citetalias{King03}, based on a different kinematic criterion\footnote{\citetalias{Montes01b} use criteria according to \citet{Eggen58,Eggen95}, which are not based on the individual components of the Galactic space motion in a similar way to \citetalias{King03}, but on both the motion projected onto the celestial sphere and the total space velocity.}. Only 7 stars are assigned to be kinematic members by both \citetalias{Montes01b} and \citetalias{King03} (see Fig.~\ref{fig:comp_kin}). The agreement between the kinematic membership lists is therefore poor, demonstrating that kinematic membership criteria alone are insufficient for determining whether a star is a member or not.

\begin{table*}
\caption[The UMa group kinematic parameters compiled from some recent studies]{\label{tab:uma_kin}The UMa group kinematic parameters compiled from some recent studies.}
\begin{tabular}{lcccccl}
\toprule
Reference&\multicolumn{2}{c}{Convergent Point}&\multicolumn{3}{c}{Space Velocity [km/s]}&Comments\\
&A (2000.0)&D (2000.0)&\multicolumn{1}{c}{$U$}&\multicolumn{1}{c}{$V$}&\multicolumn{1}{c}{$W$}\\
\midrule
\citet{Chereul99}&&&14.0$\pm$7.3&1.0$\pm$6.4&-7.8$\pm$5.5&scale 3\\
\citet{Chupina01}&$300\fdeg6864$&$-29\fdeg74465$&13.5$\pm$0.8&3.0$\pm$0.9&-7.5$\pm$2.0&UMa nucleus\\
\citet{Asiain99}&&&8.7$\pm$6.6&2.8$\pm$4.1&-6.9$\pm$5.8&group A\\
\citetalias{Montes01b}&20\fhour55&-38\fdeg10&14.9&1.0&-10.7\\
\citetalias{King03}&&&13.9$\pm$0.6&2.9$\pm$0.9&-8.4$\pm$1.3&UMa nucleus, weighted\\
\bottomrule
\end{tabular}\\
The comments in the last column specify details regarding the given literature source.\end{table*}

̵

The selection of the most recent membership list \citepalias{King03} was based on a combination of kinematic, photometric, and spectroscopic membership criteria. Out of 220 candidates, nearly 60 assured members are presented, although only the selection of 30 of these is consistent with the applied kinematic criteria. We study the homogeneity of spectroscopic UMa group features, so our conclusions are biased when using the final member list of \citetalias{King03} as it is.

Presently, the best solution seems to be to study different lists of members instead of trying to define a new ultimate list. Therefore, spectroscopic properties are considered separately for the list of kinematic members (lists {\it M01k} and {\it K03k}) and for the list of final members of \citetalias{King03} (list {\it K03f}):
\begin{itemize}
\item {\it M01k} -- 28 kinematic members according to \citetalias{Montes01b} (fulfilling both of Eggen's criteria);
\item {\it K03k} -- 36 kinematic members according to \citetalias{King03} (assignment {\em Y});
\item {\it K03f} -- 31 final members according to \citetalias{King03} (assignment {\em Y} without question mark).
\end{itemize}
From these (overlapping) lists of in total 58 stars, the 25 low-mass northern members\footnote{The 21 early-type stars in the final list are not studied in this work. Furthermore, five stars on the southern hemisphere are omitted since our observations were restricted to the northern hemisphere. Five more stars are close companions of brighter members and could not be resolved with the spectrographs.} are the subject to the present work (Table~\ref{tab:uma_mem_data}). 

Iron abundances are presented for the 17 Sun-like stars, six of which already were analysed with the same methods. For these stars and eight more, i.e., in total 25 late-type stars, Li absorption and chromospheric infilling of H$\alpha$ is addressed.

Table~\ref{tab:uma_mem_data} lists stellar data relevant to this work. As in \citetalias{Fuhrmann04}, $V$ band magnitudes are derived from {\em Hipparcos} photometry using the conversion of \citet{Bessell00}. In the case of the five stars analysed by \citetalias{Fuhrmann04}, the slightly different values given by \citetalias{Fuhrmann04} were adopted for consistency reasons. 

\begin{table*}
\caption{\label{tab:uma_mem_data}The sample of UMa group members.}
\begin{tabular}{lllrccccccccc}
\toprule
Object&Spectral Type&\multicolumn{1}{c}{$V$}&\multicolumn{1}{c}{$\pi$}      &\multicolumn{2}{c}{\citetalias{Montes01b}}    &\multicolumn{3}{c}{\citetalias{King03}}&\citetalias{Fuhrmann04}&\multicolumn{3}{c}{Sub-sample}\\
&&&\multicolumn{1}{c}{[mas]}&pec. vel.&rad. vel.&kin.&phot.&final       &&&&\\
\midrule
HD  11131          &G1V     & $6.727\pm0.005$$^1$&  42.35$\pm$0.87$^3$& Y & Y &?   &Y?&Y?  &Y&{\it M01k}&&\\
HD  24916       A  &K4V     & $8.033\pm0.018$    &  63.41$\pm$2.00& Y & Y &N?  &? &N?  & &{\it M01k}&&\\
HD  26923          &G0IV    & $6.314\pm0.005$$^1$&  47.20$\pm$1.08& Y & Y &Y?  &Y &Y?  &Y&{\it M01k}&&\\
HD  28495          &G0V     & $7.748\pm0.023$    &  36.32$\pm$1.07& N & Y &Y   &Y &Y   & &&{\it K03k}&{\it K03f}\\
HD  38393          &F7V     & $3.586\pm0.0044$   & 111.49$\pm$0.60& Y & Y &?/Y?&Y?&Y?  & &{\it M01k}&&\\
HD  39587          &G0V     & $4.396\pm0.005$$^1$& 115.43$\pm$1.08&   &   &Y   &Y &Y   &Y&&{\it K03k}&{\it K03f}\\
HD  41593          &K0V     & $6.762\pm0.005$$^1$&  64.71$\pm$0.91& Y & Y &N?/?&Y &N?/?&Y&{\it M01k}&&\\
HD  59747          &G5V     & $7.677\pm0.005$$^1$&  50.80$\pm$1.29& Y & Y &Y   &Y?&Y   &Y&{\it M01k}&{\it K03k}&{\it K03f}\\
HD  71974       A  &G5V     & $7.265\pm0.011$    &  34.83$\pm$1.37& Y & N &Y   &Y?&Y   & &&{\it K03k}&{\it K03f}\\
HD  95650          &M0      & $9.508\pm0.024$    &  85.76$\pm$1.36& N & Y &Y   &Y?&Y   & &&{\it K03k}&{\it K03f}\\
HD  238087         &K5      & $9.975\pm0.030$    &  35.24$\pm$1.24& Y & Y &Y?  &? &?   & &{\it M01k}&&\\
HD  109011         &K2V     & $8.069\pm0.015$    &  42.13$\pm$3.11&   &   &Y   &Y &Y   & &&{\it K03k}&{\it K03f}\\
HD  109647         &K0V     & $8.503\pm0.020$    &  38.08$\pm$1.11& Y & Y &Y   &Y &Y   & &{\it M01k}&{\it K03k}&{\it K03f}\\
HD  110463         &K3V     & $8.256\pm0.018$    &  43.06$\pm$0.82& Y & Y &Y   &Y &Y   & &{\it M01k}&{\it K03k}&{\it K03f}\\
HD  112196         &F8V     & $6.985\pm0.015$    &  29.19$\pm$1.60& Y & Y &N?/?&? &N?/?& &{\it M01k}&&\\
HD  115043         &G1Va    & $6.814\pm0.011$    &  38.92$\pm$0.67& Y & Y &Y   &Y &Y   & &{\it M01k}&{\it K03k}&{\it K03f}\\
HD  238224         &K5      & $9.642\pm0.022$    &  39.84$\pm$1.44& Y & Y &Y   &? &Y?  & &{\it M01k}&{\it K03k}&\\
HD  152863      A  &G5III   & $6.066\pm0.008$    &   7.14$\pm$0.67& Y & Y &?   &Y?&?   & &{\it M01k}&&\\
HD  155674      A  &K0      & $8.903\pm0.087$    &  47.14$\pm$1.88& Y & Y &?   &? &?   & &{\it M01k}&&\\
HD  155674      B  &K8      & $9.304\pm0.054$    &  47.86$\pm$3.11& Y & Y &?   &? &?   & &{\it M01k}&&\\
HD  167389         &F8      & $7.383\pm0.010$    &  29.91$\pm$0.59& Y & Y &N?  &? &N?/?& &{\it M01k}&&\\
HD  171746      A  &G2V     & $6.924\pm0.071$$^2$&  29.23$\pm$1.54& N & Y &Y   &Y?&Y   & &&{\it K03k}&{\it K03f}\\
HD  171746      B  &G2V     & $7.009\pm0.071$$^2$&  29.23$\pm$1.54&   &   &Y   &Y?&Y   & &&{\it K03k}&{\it K03f}\\
HD  184960         &F7V     & $5.714\pm0.006$    &  39.08$\pm$0.47& Y & Y &?/Y?&? &?/Y?& &{\it M01k}&&\\
HD  205435         &G8III   & $3.986\pm0.006$    &  26.20$\pm$0.51& Y & Y &?/Y?&? &?   & &{\it M01k}&&\\
\bottomrule
\end{tabular}\\
The table gives the spectral type {\bf (2)}, the V band magnitude {\bf (3)} derived from the {\em Hipparcos} system using the conversion of \citet{Bessell00} in most cases, the parallax $\pi$ {\bf (4)}, Eggen's peculiar velocity criterion {\bf (5)} and Eggen's cluster radial velocity criterion {\bf (6)} as applied by \citetalias{Montes01b}, the kinematic $UVW$ criterion {\bf (7)},  photometric criterion {\bf (8)}, and  final membership according to \citetalias{King03} {\bf (9)}, and the spectroscopic membership following \citetalias{Fuhrmann04} {\bf (10)}. The last column indicates the membership of the sub-samples {\it M01k}, {\it K03k}, and {\it K03f} as described in the text in detail. Most astrometric and photometric measurements are taken from the {\em Hipparcos} catalogue \citep{Perryman97}. Exceptions are indicated in the footnote below.\\
$^1$ \citetalias{Fuhrmann04}.\\
$^2$ Based on values in the {\em Hipparcos} system given by \citet{Fabricius00}. See Sect.~\ref{sect:mult} for more details.\\
$^3$ According to \citetalias{Fuhrmann04}, the parallax of the primary HD\,11171 is adopted.
\end{table*}

\section{Observations and data reduction}

The spectra were obtained in 2002-2004 with the high resolution
\'Echelle spectrograph FOCES at the 2.2\,m-telescope of the
Centro Astron\'omico Hispano Alem\'an (CAHA) at Calar Alto
\citep{Pfeiffer98,Grupp03,Grupp04}, and the Coud\'e-\'Echelle
spectrograph at the 2\,m-Alfred-Jensch telescope of the Th\"uringer
Landessternwarte at Tautenburg \citep{Hatzes05, Koenig03}. Since the analysis works differentially with respect to the Sun, we
also obtained solar spectra by observing the illuminated disk of the Moon
during each observing campaign\footnote{The observations of HD\,109011 are the only case that solar spectra have not been obtained during the same campaign.}.

Some of the stars were already analysed by \citetalias{Fuhrmann04}, although no measurement of either the Li equivalent width or the chromospheric flux was made. Spectra of these stars observed in the same way with FOCES as in the present work were kindly provided by K. Fuhrmann for this purpose.

The FOCES setup included a slit width of $120\,\mathrm{\mu m}$ and the use of the Loral 11i 2kx2k chip. Spectra with a resolution of
$\lambda/\Delta\lambda\gtrsim60\,000$ were obtained. The TLS \'Echelle observations were characterized by a slit width of $520\,\mathrm{\mu m}$ and the use of the 2kx2k EEV chip with 13.5x13.5\,$\mathrm{\mu m}$ pixels achieving a resolution of $\lambda/\Delta\lambda=67\,000$. A signal-to-noise ratio of at least 200 close to H$\alpha$ was obtained for most spectra.

For our analyses, we used the wavelength region from 4700 to 7400\,{\AA} in the
case of the TLS \'Echelle spectrograph, and from 4100 to 8200\,{\AA} in the
case of FOCES. Both spectrographs cover simultaneously all relevant
Fe\,I and FeII-lines\footnote{Not all lines can be used for each star. The presence of blends as well as the equivalent width of individual lines may vary strongly between the stars. Only lines with equivalent widths lower than some 80\,{m\AA} may be used since the UMa group is young and stronger lines might be affected by chromospheric effects. Usually, 30 Fe\,I and 10 Fe\,II lines out of the full list can be used for an individual star.} (Table~\ref{tab:qline_fe_used}), as well as H$\alpha$, H$\beta$, the Mg\,Ib triplet, and the Li\,6707.8\,{\AA} doublet (Table~\ref{tab:mg_li_linedata}). 

Using the FOCES EDRS data reduction package that Klaus Fuhrmann adapted for use with the TLS \'Echelle, we followed the usual steps of data reduction \citep{Horne86,McLean97}: bias subtraction, removal of scattered light,
simple order extraction, wavelength calibration using Th-Ar exposures,
flat-field division, rectification, and merging of all \'Echelle orders into one spectrum.

The accurate normalization of the relative continuum -- often called `rectification' -- is very important in the region of the Balmer lines, whose wings are required for the determination of the effective temperatures. The problem is that chromospheric lines such as H$\alpha$ are so broad that the wings of these lines extend over more than one \'Echelle-order. The normalization of the FOCES spectra is straightforward and completed by using spectra taken with the flat-field lamp, because the illumination of either a lamp or a star is identical in this fibre-fed spectrograph (see \citealp{Korn02} for a related discussion). In Tautenburg, spectra of a standard flat-field lamp placed in front of the slit was found to be unsuitable for the normalization. Thus, a new flat-field device was installed, where four 500\,W continuum lamps illuminate homogeneously a white screen in the dome at which the telescope is pointed. The light from the lamp is then reflected in the same way as starlight before entering the slit of
the spectrograph. This is the case for all mirrors of the telescope including the Coud\'e-mirror train. The flat-field device was tested thoroughly by observing many stars, including $\alpha$\,Lyr, and found to be capable for normalizing well the spectra. 

\section{Stellar parameters and iron abundance}

\subsection{Methods}

To derive the stellar parameters, model atmospheres in local thermodynamic equilibrium (LTE) were used. These comprise temperature and pressure profiles computed with {\em MAFAGS}, an unpublished code originally developed by T. Gehren and later revised by
\citet{Reile87}. Spectral profiles are calculated using the unpublished line formation code {\em LINFOR}, which is based on {\em MAFAGS} and was developed by T. Gehren, C. Reile, K. Fuhrmann, and J. Reetz.

The model atmospheres are based on the plane-parallel approximation and hydrostatic equilibrium is assumed. The elemental abundance pattern is based mostly on \citet{Holweger79}. Line opacity is accounted for by scaled opacity distribution functions \citep{Kurucz95}. Convection is accounted for by the mixing-length approximation. {\em LINFOR} reads the temperature and pressure profiles and solves the radiative transfer in LTE for hydrogen, neutral, and singly ionized iron as well as neutral magnesium. For more details, the reader is referred to \citet{Fuhrmann93a} and \citet{Fuhrmann97a}. Additional information about {\em MAFAGS} and {\em LINFOR} is given in both \citet{Ammler06a,Ammler06b} and \citet{Korn05}.

Since Fuhrmann's method for deriving the stellar parameters was extensively described in
the literature \citep{Fuhrmann93a,Fuhrmann93b,Fuhrmann97a,Fuhrmann98,Ammler06a,Ammler06b}, only the basic ideas are summarized briefly here. Effective temperature,
surface gravity, microturbulence, and iron abundance are determined iteratively in several steps. Effective temperature is derived from the
Balmer line wings, and the surface gravity is determined from either the
wings of the Mg\,Ib triplet or the ionization equilibrium of iron
(Fe\,I and Fe\,II). The iron abundance, the microturbulence parameter, and the
projected rotational velocity are determined from the profiles of
unblended Fe\,II lines.

\subsubsection{Effective temperature}

Effective temperature is derived by fitting synthetic line profiles to
the wings of the H$\alpha$\footnote{The H$\beta$ and H$\gamma$ lines are strongly blended in all of the studied spectra and do not allow one to derive effective temperature. H$\beta$ can still be used to check the consistency of the temperatures derived with H$\alpha$.} line according to \citet[][and previous
publications]{Fuhrmann94}. One minor complication is that the UMa
group is so young that the core of H$\alpha$ is partly filled-in by
chromospheric emission. To estimate the effect of the
chromospheric infilling, temperature is assessed additionally by fitting the
wings of H$\beta$, which is less affected by chromospheric activity
\citep[also see][]{Fuhrmann04}. Both values agree well leading to the conclusion that the infilling mostly affects the cores, not the
wings of the lines. Only in the case of HD\,28495 are there residuals of the fits in the inner part of the H$\alpha$ wings that may be due to activity (see discussion of individual objects in Sect. \ref{sect:mult}). This object displays an elevated level of chromospheric flux (see Table~\ref{tab:memK_spec} and Figs.~\ref{fig:teff_ha} and \ref{fig:plot_lyra}). 

An error bar is estimated from the noise, by fitting upper and lower envelopes to the noise features. These estimates are usually below $100\,$K.

\subsubsection{Surface gravity from the iron ionization equilibrium}
Surface gravity is derived either from the iron ionization
equilibrium, or from the wings of the Mg\,Ib triplet, or from the {\em Hipparcos} distance if both spectroscopic methods fail.

\label{sect:ion_equ}
For all stars with temperatures between 5000 and 6000\,K
(HD\,28495, HD\,109647, HD\,110463,
HD\,115043), the iron ionization balance was imposed to derive the surface gravity. This method requires the measurement of the equivalent widths of about 30 unblended Fe\,I lines and 10 of Fe\,II. These are inspected and selected individually for each spectrum from a list of more than 100 lines (see Table~\ref{tab:qline_fe_used}).

The derived surface gravity varies strongly when changing the adopted effective temperature within its error bars. Furthermore, the result depends on the adopted iron abundance and the value of the microturbulence parameter\footnote{Except in the case of very weak lines, the adoption of a wrong microturbulence parameter causes spurious abundance determinations. To have a sufficient number of lines at our disposal, we need to include all useful lines with equivalent widths $<80\,$m{\AA} and therefore have to account properly for microturbulence.}. Therefore, surface gravity is determined iteratively together with iron abundance and the microturbulence parameters (both from Fe\,II lines only), i.e., all of these parameters are determined consistently from the spectrum. 

Error bars for all these values are estimated by varying the effective temperature within its error bars. The resulting error bar of surface gravity is typically smaller than $0.15\,$dex. These values may underestimate the true errors involved. Since all possible systematic errors cannot be practically taken into account, the reliability of the surface gravities is assessed by various tests (Sect.~\ref{sect:test}).

\subsubsection{Surface gravity from strong lines}

\begin{table}
\caption{\label{tab:mg_li_linedata}Mg\,I and Li\,I line data.}
\begin{tabular}{cccccc}
\toprule
$\lambda$&$\chi_\mathrm{low}$&${\log}gf$&${\log}C_6$&$\gamma_\mathrm{rad}$&${\log}C_4$\\
$[\AA]$&[eV]&&&$10^8$\,[1/s]&\\
\midrule
 \multicolumn{6}{c}{Mg\,I}\\
 4571.097&  0.00& -5.59& -31.50& --         &-15.00\\
 4730.031&  4.33& -2.34& -28.80& 4.768     &-15.00\\
 5167.336&  2.70& -0.87& -30.88& 1.002     &-14.52\\
 5172.697&  2.70& -0.39& -30.88& 1.002     &-14.52\\
 5183.616&  2.70& -0.17& -30.88& 1.002     &-14.52\\
 5528.415&  4.33& -0.50& -30.43& 4.908     &-13.12\\
 5711.093&  4.33& -1.70& -30.00& 4.817     &-15.00\\
 5785.280&  5.11& -1.76& -30.20& --         &-15.00\\
 \midrule
 \multicolumn{6}{c}{Li\,I}\\
 6707.760&  0.00&  0.00& -31.86& --         &-15.00\\
 6707.910&  0.00& -0.31& -31.86& --         &-15.00\\
\bottomrule
\end{tabular}\\
Following values are listed: {\bf (1)} central wavelength, {\bf (2)} lower excitation potential, {\bf (3)} oscillator strength, {\bf (4)} van der Waals damping constant, {\bf (5)} radiative damping constant, {\bf (6)} quadratic Stark effect damping constant. The radiation damping constant ($\gamma_\mathrm{rad}$) is calculated with the classical formula within {\em LINFOR} in case it is missing in the table.
\end{table}

For stars with a temperature of about 5900\,K or more, the application of the ionization equilibrium may underestimate the surface gravity (i.e., for HD\,167389, HD\,184960). The problem seems to arise from the Fe\,I lines. In the case of the warmer stars, most iron is ionized and the Fe\,I lines become strongly dependent on the details of the temperature structure, which might be insufficiently predicted by the LTE model atmospheres \citep[cf.][]{Fuhrmann97a}. In these cases, surface gravity is assessed by fitting the wings of the Mg\,Ib triplet\footnote{For a discussion of the preference of magnesium with respect to other strong lines, we refer to \citet{Fuhrmann97a}.}. The best results are obtained with
Mg\,I\,$\lambda5172\,${\AA} and $\lambda5183\,${\AA}. An estimate of 
Mg abundance is needed and is obtained from weak Mg\,I lines
(e.g., $\lambda4571\,${\AA} and $\lambda5711\,${\AA}, see Table~\ref{tab:mg_li_linedata}). In the case of HD\,167389, the fit is complicated and may produce a slightly spurious surface gravity (see discussion of individual objects in Sect. \ref{sect:mult}). 

The error bars are inferred in the same way as before (Sect.~\ref{sect:ion_equ}) but now the iteration also incorporates the derivation of the Mg abundance. The resulting error bars are roughly the same as in the use of the iron ionization balance (also see \citealp{Fuhrmann97a}).

\subsubsection{Surface gravity from {\em Hipparcos} parallaxes}
\label{sect:hipparcos}
For some stars hotter than $\approx5900\,$K, we found that even the use of the magnesium wings fails (HD\,38393, HD\,112196, HD\,171746\,A, and HD\,171746\,B), and therefore, surface gravity is calculated from {\em
Hipparcos} parallaxes $\pi$ \citep[for details
see][]{Fuhrmann97b,Ammler06a,Ammler06b}:

\begin{eqnarray}
\label{equ:logg}
{\log}g=&2{\log}\pi+4.44+{\log}M+4{\log}\frac{T_{\mathrm{eff}}}{5780\,\mathrm{K}}\\
\nonumber
+&\frac{2}{5}(V-A_{\mathrm{V}}-B.C._\mathrm{V}+5-M_{\mathrm{bol},\odot}),
\end{eqnarray}

\noindent
where $\pi$ is the parallax.
The apparent visual magnitude $V$ is derived from the
{\em Hipparcos} $H_\mathrm{P}$ magnitudes using the conversion tables of
\citet{Bessell00}. The adopted parallaxes and magnitudes are given in Table~\ref{tab:uma_mem_data}. The bolometric corrections $B.C._\mathrm{V}$ are
taken from \citet{AAM95}\footnote{They use another definition of
$B.C._\mathrm{V}$, which has an opposite sign!}. All of the stars analysed in this work are closer than 40\,pc so that interstellar absorption is considered neglible and not accounted for ($A_\mathrm{V}=0$).
The stellar mass $M$ finally is given in units of solar mass. Since the UMa group has an age of $\approx\,10^8$\,Myrs and all sample stars have spectral types from late-FV to early-KV, mass is estimated by simply comparing the stellar parameters to the theoretical zero-age main sequence of
\citet{VandenBerg00} ($\mathrm{[Fe/H]}=0.00$,
$[\alpha/\mathrm{Fe}]=0.0$).

An error bar is obtained from Gaussian error propagation using partial derivatives (see Tables~\ref{tab:uma_mem_data} and \ref{tab:uma_mem_par} for the individual contributions entering Eq.~\ref{equ:logg}). While photometric variability would lead to larger errors, we expect that it is low \citep[cf.][]{Adelman00} for the warmer stars addressed here. The resulting gravity error bars are $0.06\,$dex or smaller.

Finally, iron abundance and microturbulence are derived from Fe\,II lines. Errors are inferred again by varying effective temperature within its error bar.

\subsubsection{Instrumental profile, macroturbulence, and rotational velocity}

The macroturbulence parameter is calculated from the linear fit given by \citet{Gray84}

\begin{equation}
\zeta_\mathrm{RT}\mathrm{[km/s]}=3.95{\cdot}T_\mathrm{eff}[10^3\mathrm{K}]-19.25
\end{equation}

The instrumental profile is obtained by convolution with the Kitt Peak solar-flux atlas \citep{Kurucz84} and comparison to the observed lunar spectrum. The projected rotational velocity $v{\sin}i$ is then inferred by fitting
individual weak Fe\,II lines. The scatter of $v{\sin}i$ is usually less than $0.50\,$km/s. However, taking into account the
uncertainties in both the instrumental broadening and the macroturbulence
parameter, the true error is probably closer to
$1.0\,$km/s.

\subsection{Accuracy of the stellar parameters}
\label{sect:test}
\begin{table*}
\caption{\label{tab:uma_mem_par}Stellar parameters of the 18 Sun-like stars..}
\begin{tabular}{lrrrrrrrrrrrrrr}
\toprule
Object&$T_\mathrm{eff}$&${\log}g$&[Fe/H]&[Mg/Fe]&\multicolumn{1}{c}{$\xi_\mathrm{t}$}&\multicolumn{1}{c}{$\zeta_\mathrm{RT}$}&\multicolumn{1}{c}{$v{\sin}i$}&ref.&\multicolumn{1}{c}{$M$}&\multicolumn{1}{c}{$R$}&$B.C._V$&$M_\mathrm{bol}$&$d_\mathrm{sp}$&\multicolumn{1}{c}{$\frac{d_\mathrm{sp}-d_\mathrm{HIP}}{d_\mathrm{HIP}}$}\\
&[K]&$[\frac{\mathrm{cm}}{\mathrm{s}^2}]$&&&[km/s]&[km/s]&[km/s]&&[$M_\odot$]&[$R_\odot$]&&&[pc]&\multicolumn{1}{c}{[\%]}\\
\toprule
\multicolumn{15}{l}{\it M01k}\\
\midrule
            HD  11131 &  5796& 4.43&-0.10& 0.00&0.97&3.70&  1.5&  2& 1.00& 1.01&-0.13& 4.73& 23.8&  0.6\\
\vspace{\tablinesep}
                                         &    70& 0.10&0.07& 0.05&0.20&& 1.0&& 0.05& 0.04&&0.07& 3.3&\\
         HD  24916\,A&  4600& 4.60&-0.03&--   &--  & -- &  2.1&  3& 0.70& 0.69&-0.57& 6.48& 15.4& -2.4\\
\vspace{\tablinesep}
                                         &   100& 0.20&0.15& --  & -- && 2.0&& 0.05& 0.03&&0.09& 3.7&\\
            HD  26923  &  5975& 4.41&-0.06&-0.01&0.99&4.40&  $<$1.0&  2& 1.07& 1.01&-0.11& 4.57& 22.4&  5.5\\
\vspace{\tablinesep}
                                         &    70& 0.10&0.07& 0.05&0.20&&--  && 0.05& 0.04&&0.07& 3.0&\\
            HD  38393    & 6310& 4.28$^1$&-0.05& 0.01&1.30&5.70&  8.1&  1& 1.23& 1.33&-0.08& 3.74&   --&  --\\
\vspace{\tablinesep}
                                         &    60& 0.03&0.05& 0.06&0.20&& 1.0&& 0.05& 0.04&&0.05&  --&\\
            HD  41593    &  5278& 4.54& 0.02&-0.01&1.12&1.70&  4.3&  2& 0.89& 0.82&-0.24& 5.57& 15.8&  2.5\\
\vspace{\tablinesep}
                                         &    80& 0.10&0.07& 0.05&0.20&& 1.0&& 0.05& 0.03&&0.06& 2.2&\\
            HD  112196    & 6110& 4.39$^1$& 0.01& 0.02&1.31&4.90& 12.0&  1& 1.15& 1.14&-0.09& 4.22&   --&  --\\
\vspace{\tablinesep}
                                        &    60& 0.06&0.05& 0.05&0.20&& 1.0&& 0.05& 0.08&&0.14&  --&\\
            HD  167389   &  5895& 4.37&-0.02&-0.03&0.99&4.00&  3.3&  1& 1.05& 1.01&-0.12& 4.64& 36.8& 10.0\\
\vspace{\tablinesep}
                                         &    80& 0.15&0.07& 0.05&0.20&& 1.0&& 0.05& 0.04&&0.07& 6.5&\\
            HD  184960   &  6310& 4.22&-0.08&-0.01&1.55&5.70&  8.0&  1& 1.23& 1.43&-0.08& 3.59& 25.6&  0.0\\
\vspace{\tablinesep}
                                        &    80& 0.10&0.05& 0.05&0.20&& 1.0&& 0.05& 0.05&&0.06& 3.1&\\
\toprule
\multicolumn{15}{l}{{\it M01k, K03k, K03f}}\\
\midrule
            HD  59747  &  5094& 4.55&-0.03&-0.01&1.00&0.90&  1.5&  2& 0.83& 0.76&-0.31& 5.90& 20.8&  5.8\\
\vspace{\tablinesep}
                                         &    90& 0.10&0.07& 0.05&0.20&& 1.0&& 0.05& 0.04&&0.08& 2.9&\\
            HD  109647   &  4980& 4.70& 0.05&-0.07&0.80&0.40&  3.1&  1& 0.83& 0.74&-0.36& 6.05& 24.0& -8.8\\
\vspace{\tablinesep}
                                          &60& 0.15&0.15& 0.15&0.20&& 1.0&& 0.05& 0.03&&0.08& 4.3&\\
            HD  110463     & 4940& 4.68& 0.00&-0.02&0.86&0.40&  3.1&  1& 0.80& 0.75&-0.38& 6.05& 20.9& -9.8\\
\vspace{\tablinesep}
                                         &    80& 0.18&0.10& 0.10&0.20&& 1.0&& 0.05& 0.03&&0.07& 4.5&\\
\vspace{\tablinesep}
            HD  115043   &  5850& 4.36&-0.10& 0.05&1.26&3.90&  7.5&  1& 1.05& 1.03&-0.13& 4.64& 28.3& 9.3\\
\vspace{\tablinesep}
                                        &    70& 0.13&0.07& 0.05&0.20&& 1.0&& 0.05& 0.04&&0.06& 4.4&\\
\toprule
\multicolumn{15}{l}{{\it K03k, K03f}}\\
\midrule
            HD  28495    & 5465& 4.45&-0.05& 0.00&0.93&2.34&  4.5&  1& 0.95& 0.85&-0.19& 5.35& 31.3& 13.8\\
\vspace{\tablinesep}
                                          &    90& 0.12&0.05& 0.05&0.20&& 1.0&& 0.10& 0.04&&0.08& 4.8&\\
            HD  39587  &  5920& 4.39&-0.07&-0.01&0.95&4.20&  8.7&  2& 1.04& 1.03&-0.12& 4.59&  9.1&  5.3\\
\vspace{\tablinesep}
                                          & 70& 0.10&0.07& 0.05&0.20&& 0.8&& 0.05& 0.04&&0.05& 1.3&\\
            HD  71974\,A$^2$   &  5570& 4.49&--&--&--&2.8&3.2&1 &--&--&--&--& --&--\\
\vspace{\tablinesep}
                                         &    100& 0.19&--&--&--&&1.0&&--&--&&--&--&\\
            HD  109011   &  5000& 4.60& -0.12&-0.03&$\lesssim0.52$&0.50&  5.0&  1& 0.83& 0.81&-0.34& 5.85& 22.3& -6.0\\
\vspace{\tablinesep}
                                          &100& 0.20&0.07& 0.07&--&& 1.0&& 0.05& 0.07&&0.17& 5.3&\\
            HD  171746\,A & 6100& 4.36$^1$&-0.04& 0.03&1.50&4.80&  8.3&  1& 1.15& 1.18&-0.10& 4.16&   --&  --\\
\vspace{\tablinesep}
                                          &    80& 0.06&0.06& 0.05&0.20&& 1.0&& 0.05& 0.08&&0.14&  --&\\
            HD  171746\,B &  6040& 4.37$^1$& 0.02&-0.04&1.51&4.60&  7.3&  1& 1.10& 1.16&-0.10& 4.24&   --&  --\\
\vspace{\tablinesep}
                                       &    80& 0.06&0.05& 0.05&0.20&& 1.0&& 0.05& 0.08&&0.14&  --&\\
\bottomrule
\end{tabular}\\
The stellar parameters of the sample stars which were studied following the methods of \citet{Fuhrmann98}. The stars are grouped according to the samples they form part of (Table~\ref{tab:uma_mem_data}). The following quantities are listed with the second line providing the errors bars: {\bf (2)} effective temperature, {\bf (3)} surface gravity, {\bf (4)} iron abundance, {\bf (5)} magnesium abundance relative to iron, {\bf (6)} microturbulence parameter, {\bf (7)} macroturbulence parameter, {\bf (8)} projected rotational velocity, {\bf (9)} source of the values: 1=this work, 2=\citetalias{Fuhrmann04}, 3=\citet{Koenig06}, {\bf (10)} mass, {\bf (11)} radius, {\bf (12)} bolometric correction based on \citet{AAM95}, {\bf (13)} absolute bolometric magnitude (from {\em Hipparcos} distance), {\bf (14)} spectroscopic distance, and {\bf (15)} relative offset from {\em Hipparcos} distance.\\
$^1$ Surface gravity derived from {\em Hipparcos} parallax as described in Sect.~\ref{sect:hipparcos}. The comparison of spectroscopic distance with {\em Hipparcos} distance cannot be done.\\
$^2$ The stellar parameters are preliminary because HD\,71974 is a spectroscopically unresolved binary (see discussion of individual objects in Sect.~\ref{sect:mult}). Abundances could not be derived.
\end{table*}

All results are listed in Table~\ref{tab:uma_mem_par}. Although error bars are estimated for all parameters, accounting for all possible systematic and random errors is hardly feasible. Therefore, accuracy is judged by several tests. The first test is a consistency test. Since the method works differentially with respect to the Sun, the analysis of the observed solar spectra has to -- and does -- reproduce the parameters of the Sun \citep[see][for details]{Ammler06a,Ammler06b}. The second test is simply to analyse a star that has already been studied previously by \citetalias{Fuhrmann04}, which in our work is HD\,217813. This star is well suited to this comparison since it is a probable UMa group member. The values that we derived are consistent with those previously found (see \citealp{Ammler05a}).

\begin{figure*}
\centering
\includegraphics[width=12cm]{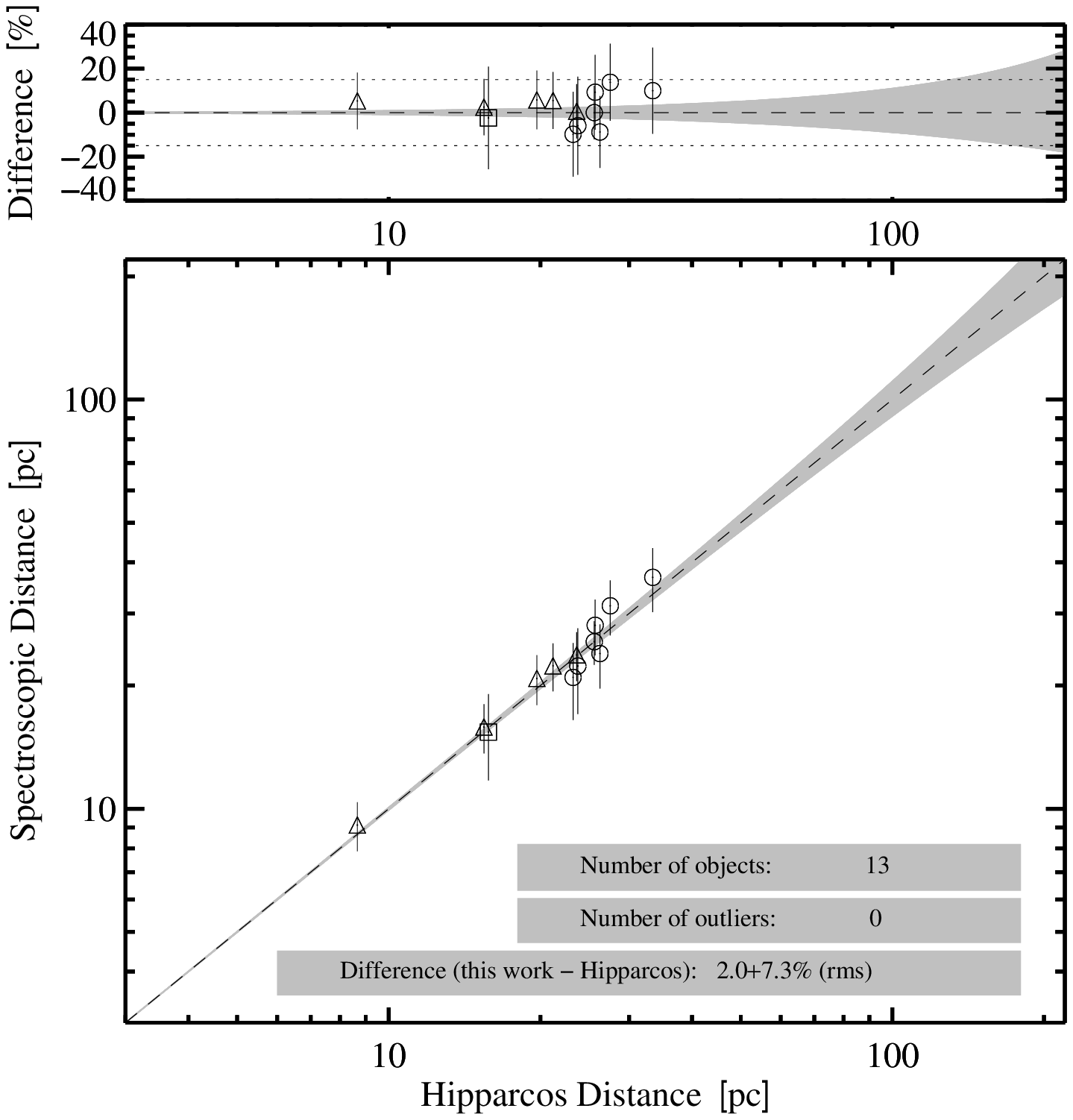}
\caption[Comparison of spectroscopic distances with {\em Hipparcos}
distances]{\label{fig:spp_aam}{\bf Comparison of spectroscopic
distances with {\em Hipparcos} distances} -- The bottom panel allows one to directly compare the distances. The dashed line represents perfect
agreement with {\em Hipparcos}. The grey-shaded area displays the {\em
Hipparcos} errors. The top panel shows the residuals and allows one to
judge the overall accuracy of the spectroscopic distances. The dotted
line represents the typical individual uncertainties of $10-15\,$\%
experienced by \citetalias{Fuhrmann04}. Data from our work are depicted by
circles, whereas the data from \citetalias{Fuhrmann04} are denoted by
triangles. The square represents the spectroscopic distance of
HD\,24916\,A, which was calculated from the parameters given by
\citet{Koenig06}.}
\end{figure*}

A crucial test of the spectroscopically derived surface gravities is the comparison of the spectroscopic distance with the {\em Hipparcos} distance (Fig.~\ref{fig:spp_aam}). Spectroscopic distance is
calculated by solving Eq.~\ref{equ:logg} for the parallax using the stellar parameters derived (Table~\ref{tab:uma_mem_par}), and the remaining quantities as described in Sect.~\ref{sect:hipparcos}. Of course
HD\,38393, HD\,112196, HD\,171746\,A, and HD\,171746\,B are not
included in this comparison, because their surface gravities were
calculated from the {\em Hipparcos} distances.

The error bars are calculated by Gaussian error propagation using the partial derivatives of Eq.~\ref{equ:logg}. At first glance, the spectroscopic distances agree very well with the {\em Hipparcos} distances within the error bars, but a closer look identifies subtle differences, which are interesting to discuss. First of all, the scatter of the new measurements is noticeably larger than that in \citetalias{Fuhrmann04} and \citet{Koenig06}. However, this is not surprising since the methods are applied here to stars that are cooler, as well as hotter, than those in these previous studies. The presently studied stars are at the edge of the temperature regime where the applied methods can be used. At very cool temperatures, the number of Fe\,II-lines becomes small\footnote{The few available Fe\,II lines were checked for blends using the {\em extract stellar} feature of {\em VALD} (\underline{V}ienna \underline{A}strophysical \underline{L}ine \underline{D}ata-Base)\index{VALD} \citealp{Kupka00,Kupka99,Ryabchikova97,Piskunov95}) with a sensitivity limit of 0.05.} and {\em MAFAGS} models of cool metal-rich atmospheres are unavailable. In the case of the hot stars, the determination of
the surface gravities from the ionization equilibrium of the Fe\,I and Fe\,II lines is complicated. It is thus unsurprising that stars with
very low, or very high temperatures are less well modelled. Another
point that should be kept in mind is binarity. A binary, in which both stars
have the same brightness, would have a greater spectroscopic
distance by up to 41\,\%. Since the luminosity of the object is needed to derive the spectroscopic distance, the distance of a binary
will always be found to be too large. A close inspection of
Fig.~\ref{fig:spp_aam} indicates that there are three stars for which the
spectroscopic distance is a bit too large, and these objects are discussed in more detail in the next section. Apart from
these subtle differences, the agreement between the spectroscopic
distance and the {\em Hipparcos} distance is fairly good so
that the stellar parameters derived in the present work are considered to be correct.

\subsection{The binary nature of individual objects}
\label{sect:mult}
The confusion between single and binary stars may produce spurious stellar parameters. In the following, the possible influences of multiplicity on the spectroscopic analyses of HD 28495, HD 115043, and HD 167389 are discussed, stars which all have a somewhat too large spectroscopic distance in Fig.~\ref{fig:spp_aam}, as well as the visual binary HD 171746 and the unresolved binary HD\,71974\,A.

\begin{itemize}

\item{{\bf HD 28495:} Until now, this star has always been referred to as
a single star. However, the catalogue of \citet{Makarov05} lists a difference of $18.3\pm2.3\,$mas/yr in the proper motion measurements of {\em Hipparcos} and {\em Tycho-2}, and an acceleration of the proper motion of $17.8\pm2.8\,\mathrm{mas}/\mathrm{yr}^2$. Following the recipe given in \citet{Makarov05} and
assuming a circular orbit, an upper limit to the mass of a
hypothetical companion of $0.20\,M_\odot$ was derived. This object would have an absolute V-band magnitude $M_\mathrm{V}=12$, and would thus be 6.5\,mag fainter than the primary \citep[Fig.~3]{BCAH98}. This faint companion would
certainly not spoil the analysis presented in the present work. It is thus more plausible that the analysis -- in particular the temperature determination -- was slightly affected by the partly infilled H$\alpha$ profile.}

\item{{\bf HD\,71974\,A:} The resulting effective temperature and surface gravity may be spurious, since the object is a close binary that cannot be resolved with spectroscopy. \citet{Fabricius00} give a separation of 0\farcs48 and a brightness difference of $\Delta{H_\mathrm{P}}=1.6$, while \citet{Heintz88} find a separation of 0\farcs377 and a brightness difference of 1.4\footnote{photometric band not given}.}

\item{{\bf HD 115043:} \citet{Lowrance05} detected a companion of
spectral type M4-M5 at a separation of $1\farcs58$. Given this spectral
type, the absolute V band magnitude of the companion is
$M_\mathrm{V}\sim11.8$ \citep[Table~3]{SK82}, compared to
$M_\mathrm{V}=4.63$ of the primary. Again, ths faint companion would not have affected the spectroscopic analysis.}

\item{{\bf HD 167389:} According to all sources listed in SIMBAD, this
object is a single star. However, the fits to the Mg\,Ib lines are insufficient possibly leading to an underestimation of the surface gravity.}

\item{{\bf HD\,171746\,A \& B:} The two components of this binary have a
separation of $1\farcs718$. The position angle is 
$158\fdeg8$ \citep{Fabricius00}. Using the photometry from \citet{Fabricius00} and converting the {\em Hipparcos} system to the Johnson
V-system \citep{Bessell00}, the A and B components have a brightness of
$V_\mathrm{A}=6.924\pm0.071$ and $V_\mathrm{B}=7.009\pm0.071$. Because of the small separation and the nearly equal brightness of the two stars, some light from the B component may have entered the fibre, when the A component was observed, and of course light from the A component may have entered when the B component was observed. Thus, the resulting stellar parameters should be regarded with care.}

\end{itemize}

\subsection{Comparison of stellar parameters with previous studies}
\label{sect:comp_prev}

Since a thorough comparison of our work with results obtained by other authors has already been presented in \citet{Ammler06a,Ammler06b}, 
only the most meaningful findings are discussed here. The comparison with \citet{KS05} is of particular interest, because they have several stars in common with the present work. Furthermore, \citet{KS05} used LTE model atmospheres, and their method is also differential with respect to the Sun. However, their derivation of stellar parameters differed from our work. They determined effective temperatures from a colour-based calibration \citep{Steinhauer03} and used surface gravities derived from evolutionary models of \citet{Yi01}. Finally, \citet{KS05} adopted the microturbulence parameter from the relation given in \citet{Allende04}.

\begin{figure*}
\centering
\includegraphics[width=12cm]{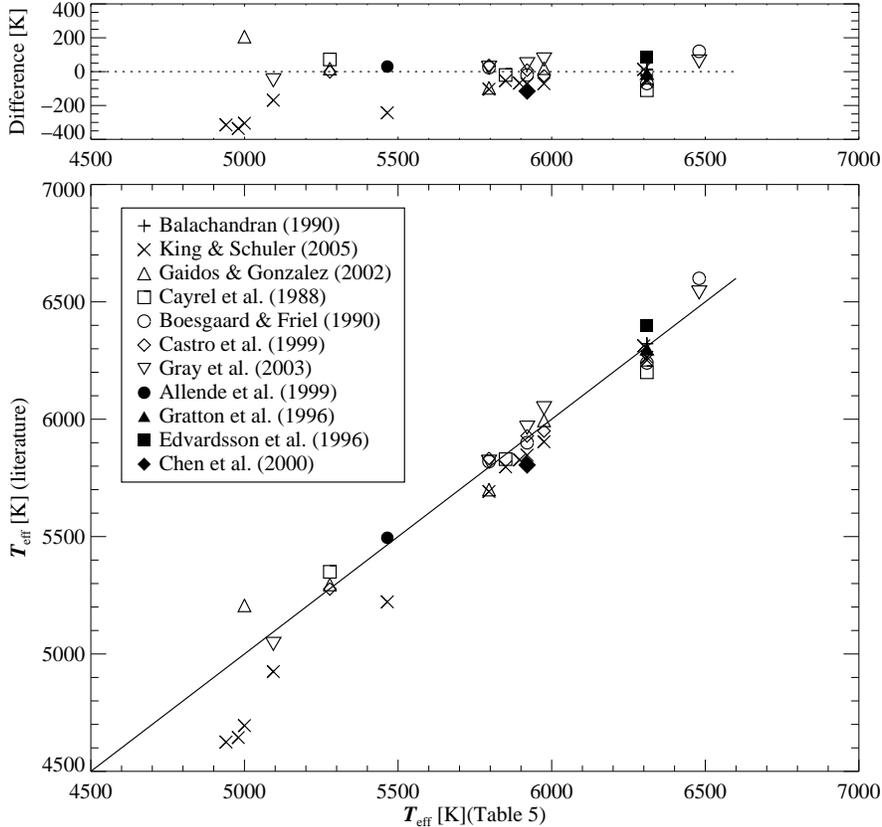}
\caption[Comparison with effective temperatures from previous analyses]{\label{fig:comp_Teff}{\bf Comparison with effective temperatures from previous analyses} -- the solid line indicates equality of {\bf the literature values and} the effective temperatures presented in Table~\ref{tab:uma_mem_par}. HD\,111456 and HD\,134083  are added for completeness. They are fast-rotating UMa group members with effective temperature estimates given by \citet{Fuhrmann00} ($T_\mathrm{eff}=6300\,$K and $T_\mathrm{eff}=6480\,$K, resp.). Error bars are omitted for clarity.
\nocite{Balachandran90}
\nocite{KS05}
\nocite{GG02}
\nocite{Cayrel88}
\nocite{BF90}
\nocite{Castro99}
\nocite{Gray03}
\nocite{Allende99}
\nocite{Gratton96}
\nocite{Edvardsson93}
\nocite{Chen00}
}
\end{figure*}

Figure~\ref{fig:comp_Teff} shows a comparison of the effective
temperatures given in Table~\ref{tab:uma_mem_par} with those derived in previous studies. All values -- except those of \citet{KS05} -- scatter closely around the solid line, which indicates equality. This and the good agreement of spectroscopic distance with the geometric {\it Hipparcos} distance generally provides confidence in the temperatures derived in the present work. A systematic discrepancy with the values derived by \citet{KS05} can be noted in Fig.~\ref{fig:comp_Teff}. This difference increases towards the lowest effective temperatures and is probably caused by the adoption of an external calibration by \citet{KS05}.

\begin{figure}
\centering
\subfigure{\includegraphics[width=8cm]{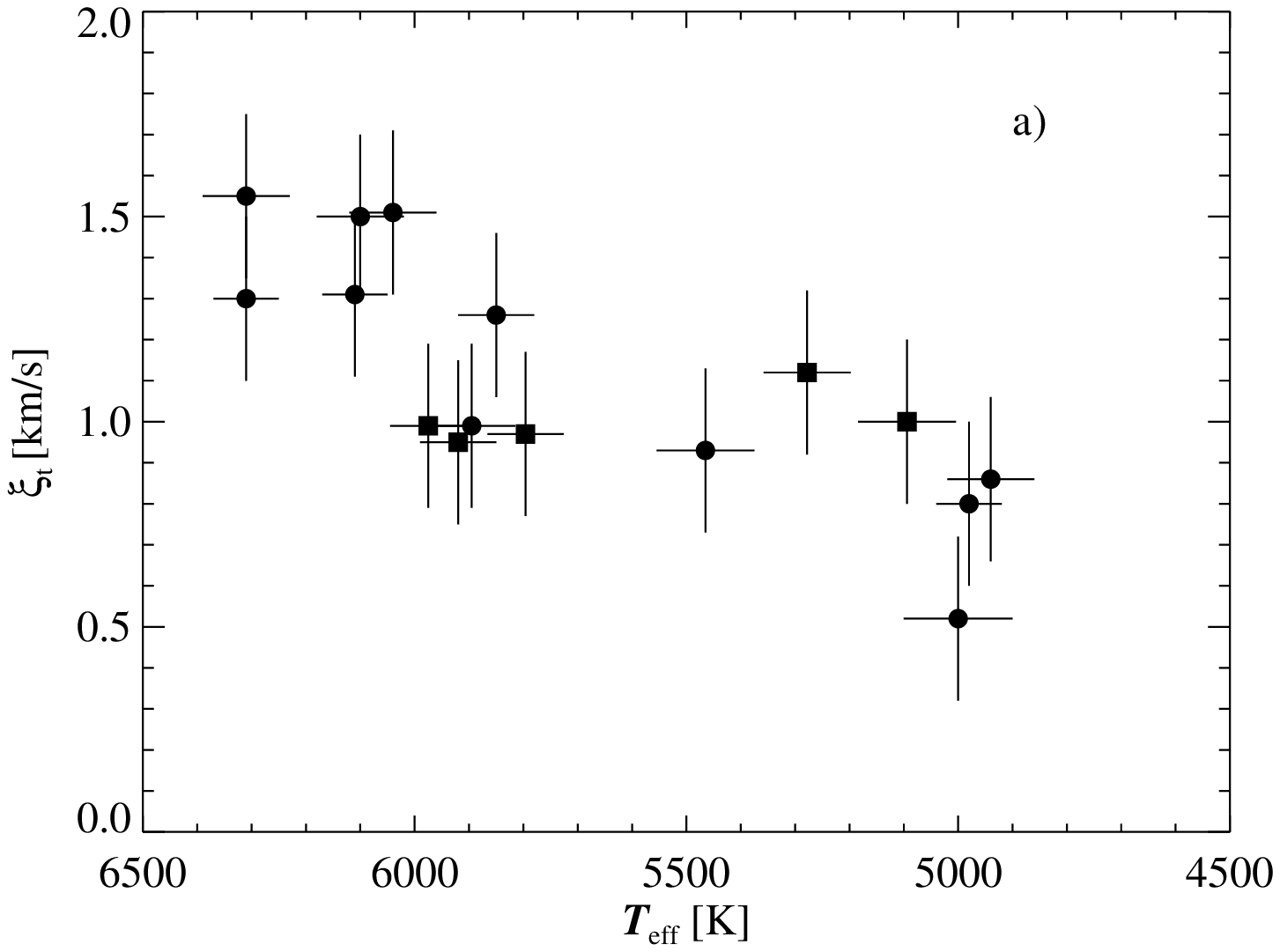}}
\subfigure{\includegraphics[width=8cm]{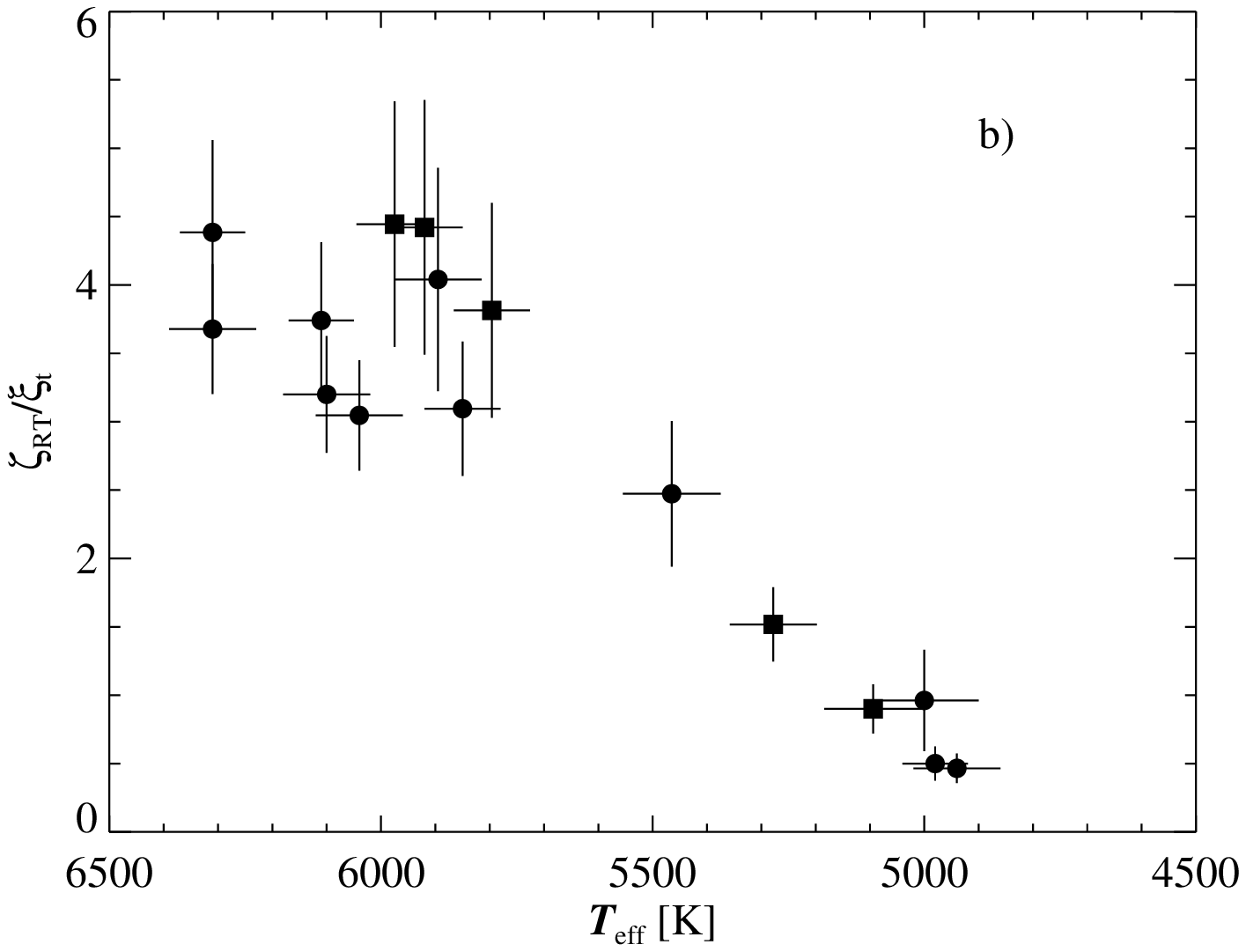}}
\caption[Comparison of the microturbulence parameter $\xi_\mathrm{t}$
with the values given by \citetalias{Fuhrmann04}]{\label{fig:comp_xi}{\bf Comparison of the
microturbulence parameter $\xi_\mathrm{t}$ with \citetalias{Fuhrmann04}} --
 all values are taken from Table~\ref{tab:uma_mem_par}. -- {\bf (a)} The values of the microturbulence parameter derived in the present work (circles) generally follow the trend with effective temperature of the values determined by \citetalias{Fuhrmann04} (squares). -- {\bf (b)}
 $\zeta_\mathrm{RT}/\xi_\mathrm{t}$ is displayed vs. efffective temperature. Symbols are the same as in the upper panel.}
\end{figure}

The abundances derived depend strongly on the value of the microturbulence parameter used within the synthetic line formation. The iron abundance is underestimated if too large a value of the microturbulence parameter is adopted. Similar to previous studies \citep[see][]{Fuhrmann04,Luck05}, an increase in the microturbulence parameter with increasing effective temperatures is found (Fig.~\ref{fig:comp_xi}). In the present case, the increase is yet higher because of the hotter stars in the sample. Since the exact value of the microturbulence parameter is determined by the way in which the analysis was done, the only meaningful comparison is that with the work of \citetalias{Fuhrmann04}, as illustrated by the different values quoted for the Sun. \citet{Fuhrmann98} gives $0.90\pm0.15$km/s, whereas \citet{Allende04} adopts $1.25\,$km/s. In the present work, solar spectra of different observing runs are analysed and an average of 0.88$\pm$0.06\,km/s is obtained, which has to and indeed does agree excellently with \citetalias{Fuhrmann04}. The lower panel of Fig.~\ref{fig:comp_xi} displays the ratio $\frac{\zeta_\mathrm{RT}}{\xi_\mathrm{t}}$ versus effective temperature, which can be compared to \citet[fig.~5]{Fuhrmann04}. The systematic difference at hotter temperatures cannot be fully explained. While somewhat higher values of the microturbulence parameter are expected in this temperature regime \citep{Fuhrmann04,GG02}, the values found in the present work (circles) follow the trend of old stars in the lower panel of Fig.~\ref{fig:comp_xi} whereas the data adopted from \citetalias{Fuhrmann04} (squares) trace the trend of the young stars according to their Fig.~5.

\section{Spectroscopic properties of the Ursa Major group}
\subsection{Distribution of iron and magnesium abundance}
\label{sect:abundances}

Table~\ref{tab:uma_mem_par} implies an average iron abundance of
$\langle[\mathrm{Fe}/\mathrm{H}]\rangle=-0.03\pm0.05$\,dex, for the kinematic members identified by \citetalias{Montes01b} (sample {\it M01K}). The value for both the kinematic ({\it K03k}) and final members ({\it K03f}) of \citetalias{King03} is $\langle[\mathrm{Fe}/\mathrm{H}]\rangle=-0.04\pm0.05$\,dex because of both the kinematic and final members of \citetalias{King03} in Table~\ref{tab:uma_mem_par} are identical. Therefore, the mean abundances of all three samples {\it M01k}, {\it K03k}, and {\it K03f} are almost the same making the definition of a mean UMa group iron abundance very robust. 

The values are slightly higher than previous measurements  ($-0.08\pm0.09$, \citealp{SM93}; $-0.09\pm0.02$, \citealp{BF90}). The abundance measurements of \citet{KS05} for stars in common imply a mean of $-0.06\pm0.08$. \citet{KS05} already highlighted the small abundance scatter of the UMa group. The spread measured in the present work is even smaller than the scatter found by \citet{KS05}. Even more, it is within the error bars of individual measurements, i.e., the iron abundances of the UMa group members are almost indistinguishable.

It is interesting to compare the iron abundances derived for the UMa group with other young stellar clusters. The Hyades are metal-rich according to \citet{BB88} ([Fe/H] = +0.17$\pm$0.06), \citet{Cayrel85} ([Fe/H] = +0.12$\pm$0.03), \citet{Paulson03} ([Fe/H] = +0.13$\pm$0.01), and
\citet{BF90} ([Fe/H]$=+0.13\pm0.05$). \citet{BF90} derive a value of
[Fe/H]$=-0.03\pm0.02$ for the Pleiades. After removing known non-members from their sample \citep{An07}, the metallicity is found to be higher, [Fe/H]$=+0.03$. This slightly super-solar value is also supported by the results of \citet{King00} ([Fe/H]$=+0.06\pm0.05$) and \citet{Grupp04} ([Fe/H]$=+0.04\pm0.02$). All in all, the iron abundance of the UMa group is about the same as that of the Pleiades, and both have about the solar abundance. Taking into account magnesium (Table~\ref{tab:uma_mem_par}), the distribution of chemical abundances coincides well with the thin disk population (\citealp[according to][Fig.~34]{Fuhrmann04}). The scatter of the UMa group Fe abundance is also at most $10\,\%$ of the total thin disk Fe abundance spread.


In an exclusive sense, a membership criterion can be obtained from the measured mean, defining the range of allowed values by $2\,\sigma$. Thus, an Fe abundance of [Fe/H]$=-0.04\pm0.10$ is expected for any UMa group member.




\subsection{Youth and activity}

\subsubsection{Lithium absorption}

\begin{table*}
\caption{\label{tab:memK_spec}Li absorption and H$\alpha$ infilling.}
\begin{tabular}{lrrrrrrrrrr}
\toprule
Object&$T_\mathrm{eff}$&$W^\mathrm{int}_\mathrm{Li}$&$W^\mathrm{fit}_\mathrm{Li}$&$W^\mathrm{KS05}_\mathrm{Li}$&${\log}N^\mathrm{cog}_\mathrm{LTE}(\mathrm{Li})$&${\log}N^\mathrm{cog}_\mathrm{NLTE}(\mathrm{Li})$&${\log}N^\mathrm{fit}_\mathrm{LTE}(\mathrm{Li})$&${\log}N^\mathrm{KS05}_\mathrm{LTE}(\mathrm{Li})$&\multicolumn{1}{c}{H$\alpha$ core}&\multicolumn{1}{c}{$F_\mathrm{H\alpha}$ [$10^5$\,erg}\\
&[K]&[m{\AA}]&[m{\AA}]&[m{\AA}]&&&&&\multicolumn{1}{c}{[\%]}&cm$^{-2}$\,s$^-1$]\\
\toprule
\multicolumn{8}{l}{\it M01k}\\
\midrule
\object{HD 11131}       & 5796$^c$&   70.5$\pm$ 1.9& 71.2&78.0&2.47$\pm$0.06&2.46$\pm$0.06&2.58$\pm$0.10&2.38$\pm$0.04&29$^c$&10.87\\
\object{HD 24916 A}     & 4600$^b$& $<$8.0$\pm$ 3.0&     &    &             &             &             &             &33    &1.08\\
\object{HD 26923}       & 5975$^c$&   88.0$\pm$ 2.1& 88.9&84.3&2.77$\pm$0.07&2.72$\pm$0.06&2.83$\pm$0.09&2.69$\pm$0.03&25$^c$&9.02\\
\object{HD 38393}       & 6310&       65.3$\pm$ 0.7& 63.5&58.5&2.87$\pm$0.05&2.81$\pm$0.04&2.96$\pm$0.09&2.82$\pm$0.03&17    &0.33\\
\object{HD 41593}       & 5278$^c$&   14.0$\pm$ 3.9& 15.8&    &1.17$\pm$0.16&1.29$\pm$0.15&1.20$\pm$0.12&             &30$^c$&8.72\\
\object{HD 238087}      & 4350$^g$&$<$20.0$\pm$ 5.0&     &    &             &             &             &             &39&--\\
\object{HD 112196}      & 6110&      109.8$\pm$ 3.1&109.5&    &3.04$\pm$0.06&2.94$\pm$0.05&3.09$\pm$0.09&             &28&11.17\\
\object{HD 152863 A}$^a$& 4980$^e$&$<$ 7.0$\pm$ 5.0&     &    &             &             &             &             &17&-0.37\\
\object{HD 155674 A}    & $\lesssim$4800 &$<$ 9.0$\pm$ 6.0&     &    &             &             &             &             &37&11.87\\
\object{HD 155674 B}    & 3900$^f$&$<$10.0$\pm$ 2.0&     &    &             &             &             &             &39&--\\
\object{HD 167389}      & 5895&       59.7$\pm$ 2.4& 59.2&58.8&2.48$\pm$0.08&2.46$\pm$0.07&2.50$\pm$0.10&2.41$\pm$0.04&21&4.70\\
\object{HD 184960}      & 6310&       60.5$\pm$ 1.9& 60.0&61.6&2.82$\pm$0.04&2.76$\pm$0.04&2.90$\pm$0.08&2.78$\pm$0.03&17&-0.02\\
\object{HD 205435}$^a$  & 5060$^g$&$<$15.0$\pm$ 5.0&     &    &             &             &             &             &20&-0.73\\
\toprule
\multicolumn{8}{l}{\it M01k, K03k}\\
\midrule
\object{HD 238224}      & 4350$^f$&$<$11.0$\pm$ 4.0&     &    &             &             &             &             &61&--\\
\toprule
\multicolumn{8}{l}{\it M01k, K03k, K03f}\\
\midrule
\object{HD 59747}       & 5094$^c$&   35.8$\pm$ 4.4& 38  &41.0&1.39$\pm$0.12&1.53$\pm$0.10&1.44$\pm$0.16&1.16$\pm$0.07&32$^c$&9.90\\
\object{HD 109647}      & 4980&       46.0$\pm$ 5.0& 44.2&25.0&1.37$\pm$0.09&1.53$\pm$0.08&1.36$\pm$0.11&0.49$\pm$0.15&34&11.65\\
\object{HD 110463}      & 4940&       21.8$\pm$ 2.6& 23.7&18.0&0.99$\pm$0.11&1.17$\pm$0.10&1.05$\pm$0.11&0.31$\pm$0.06&32&10.03\\
\object{HD 115043}      & 5850&       94.0$\pm$ 6.0& 84.0&101&2.70$\pm$0.08&2.66$\pm$0.06&2.78$\pm$0.11&2.61$\pm$0.08&27&9.48\\
\toprule
\multicolumn{8}{l}{\it K03k, K03f}\\
\midrule
\object{HD 28495}       & 5465&       69.0$\pm$ 2.4& 72.8&71.5&2.14$\pm$0.09&2.18$\pm$0.07&2.21$\pm$0.08&1.86$\pm$0.05&42&20.43\\
\object{HD 39587}       & 5920$^c$&  103.6$\pm$ 4.4&104.7&102.8&2.83$\pm$0.08&2.77$\pm$0.07&2.91$\pm$0.10&2.74$\pm$0.02&28$^c$&11.14\\
\object{HD 71974 A}     & 5570&       68.0$\pm$ 4.0& 66.7&    &             &             &             &             &32&12.90\\
\object{HD 95650}       & 3634$^d$&$<$45.0$\pm$ 5.0&     &    &             &             &             &             &73&--\\
\object{HD 109011}      & 5000&       29.0$\pm$ 6.0& 25.9&23.4&1.19$\pm$0.16&1.35$\pm$0.14&1.22$\pm$0.20&0.55$\pm$0.04&36&12.54\\ 
\object{HD 171746 A}    & 6100&       84.9$\pm$ 3.5& 83.3&    &2.85$\pm$0.08&2.79$\pm$0.07&2.90$\pm$0.10&             &17&10.34\\
\object{HD 171746 B}    & 6040&       50.0$\pm$ 5.0& 45.3&    &2.51$\pm$0.09&2.48$\pm$0.08&2.51$\pm$0.11&             &27&7.26\\
\bottomrule
\end{tabular}
\newline
{\bf (2)} Effective temperature is adopted from Table~\ref{tab:uma_mem_par} if available, or from other sources as indicated in the footnotes to the table. -- {\bf (3)} Equivalent width $W^{\mathrm{int}}_\mathrm{Li}$ of Li\,I\,${\lambda}6707.8{\AA}$ from direct integration of the observed profile. -- {\bf (4)} Equivalent width $W^{\mathrm{fit}}_\mathrm{Li}$ from fitting with synthetic line profiles. -- {\bf (5)} Equivalent width $W^{\mathrm{KS05}}_\mathrm{Li}$ according to \citet{KS05}. -- {\bf (6)} Li abundance ${\log}N^\mathrm{cog}_\mathrm{LTE}(\mathrm{Li})$ using $W^{\mathrm{int}}_\mathrm{Li}$ by interpolation of the curves of growth of \citet{Soderblom93a} -- {\bf (7)} after application of the non-LTE corrections of \citet{Carlsson94} (${\log}N^\mathrm{cog}_\mathrm{NLTE}(\mathrm{Li})$). -- {\bf(8)} Li abundance ${\log}N^\mathrm{fit}_\mathrm{LTE}(\mathrm{Li})$ derived by fitting a synthetic profile to the observed feature. -- {\bf (9)} Li abundance ${\log}N^\mathrm{KS05}_\mathrm{LTE}(\mathrm{Li})$ according to \citet{KS05}. -- {\bf (10)} Relative infilling of the core of H$\alpha$ measured according to \citetalias{Fuhrmann04}. -- {\bf (11)} Chromospheric flux $F_\mathrm{H\alpha}$ measured according to \citet{Lyra05}.\\$^a$\footnotesize{giant.}
$^b$\footnotesize{from \citet{Koenig06}.}
$^c$\footnotesize{from \citetalias{Fuhrmann04}.}
$^d$\footnotesize{from \citet{Alonso96}.}
$^e$\footnotesize{from \citet{deLaverny03}.}
$^f$\footnotesize{converted from spectral type with the calibration from \citet{KH95}.}
$^g$\footnotesize{from \citet{McWilliam90}.}
\end{table*}

\begin{figure*}
\centering
\subfigure{\includegraphics[width=\columnwidth]{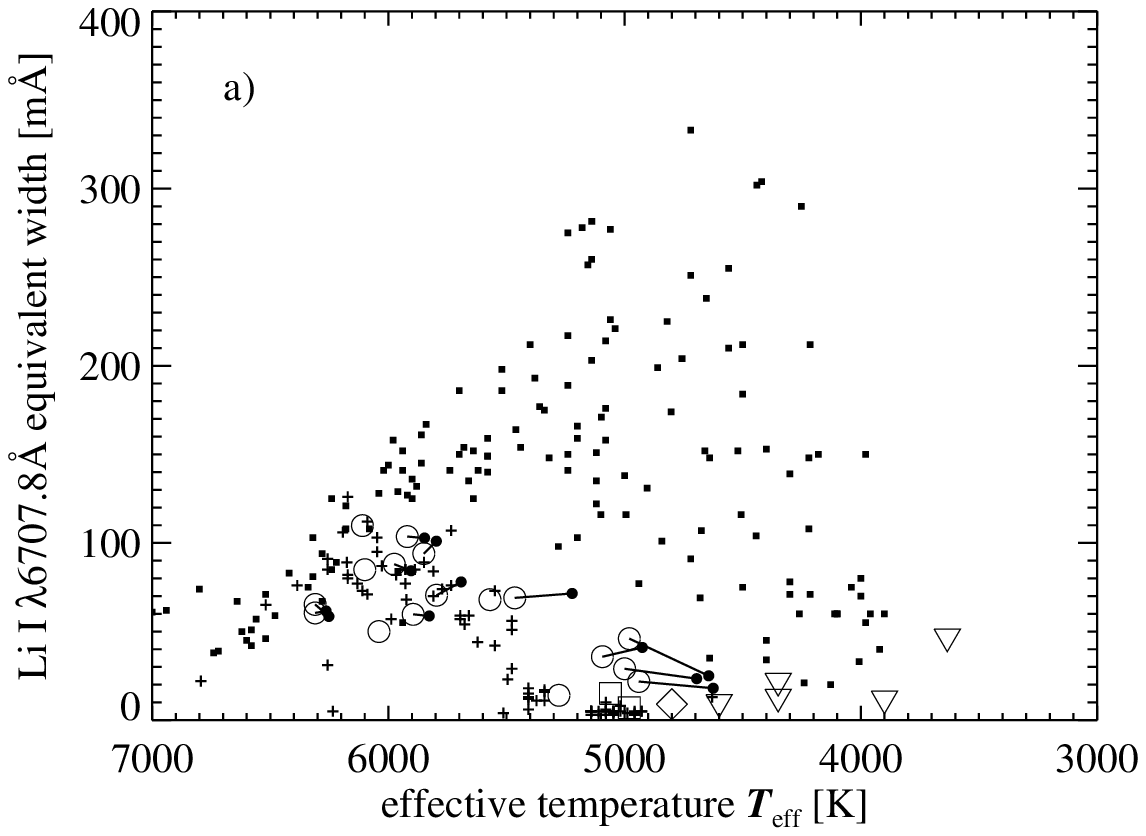}}
\subfigure{\includegraphics[width=\columnwidth]{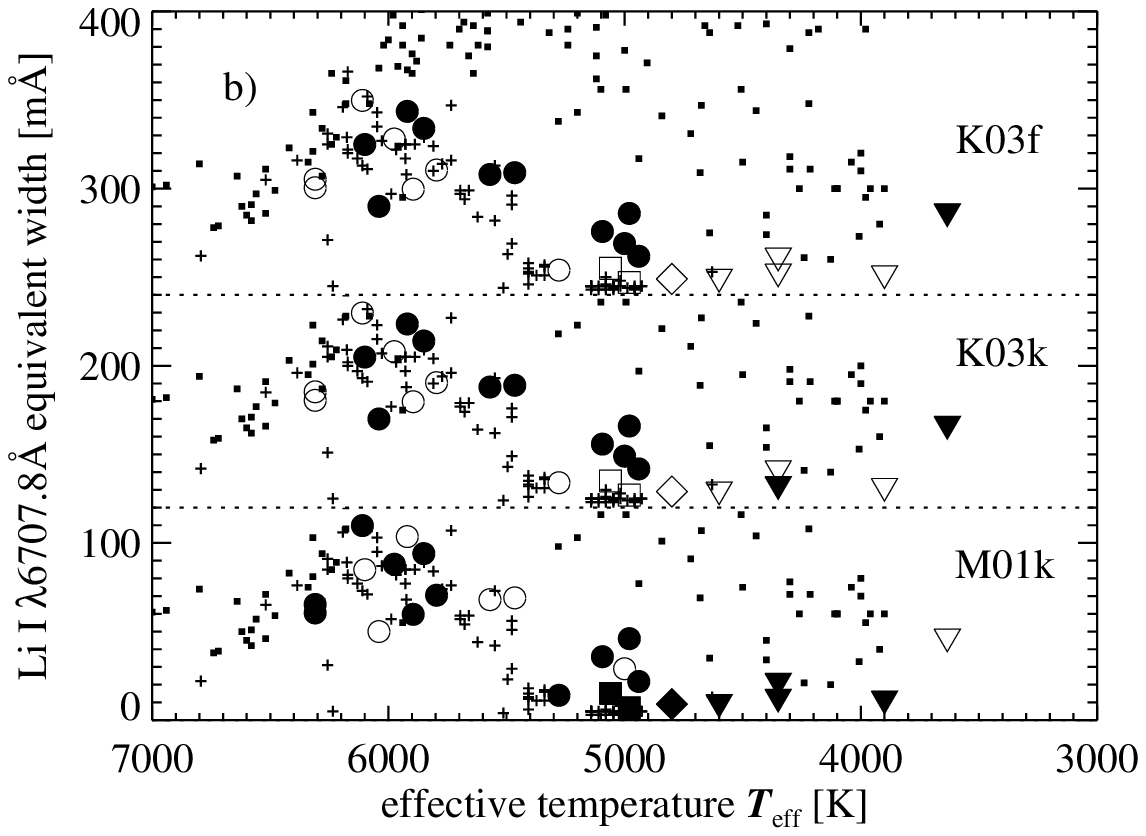}}
\caption{\label{fig:teff_ewli}{\bf Effective temperature vs. lithium absorption} -- 
{\bf a)} The uniform measurements of this work (open symbols, Table~\ref{tab:memK_spec}, $W^\mathrm{int}_\mathrm{Li}$) are compared to previous determinations by \citet{KS05} for all stars in common (dots, connected by lines, Table~\ref{tab:memK_spec}, $W^{\mathrm{KS05}}_\mathrm{Li}$). Furthermore, the distribution of the Pleiades (filled squares) and Hyades (crosses) is shown by adopting the data from \citet{Sestito05b}. The measurements of this work are depicted by circles, while upper limits are given by triangles and in the cases of the two giants HD\,152863\,A and HD 205435 by squares. In the case of HD\,155674\,A, the diamond depicts upper limits both to Li equivalent widths and effective temperatures. Error bars are omitted for clarity. -- {\bf b)} The distributions are repeated for the different samples {\it M01k}, {\it K03k}, and {\it K03f} according to Table~\ref{tab:uma_mem_data}, using arbitrary offsets. The dotted lines indicate an equivalent width of zero  to aid the eye. The members of each sample are depicted by filled symbols.}
\end{figure*}

\begin{figure}
\centering
\includegraphics[width=\columnwidth]{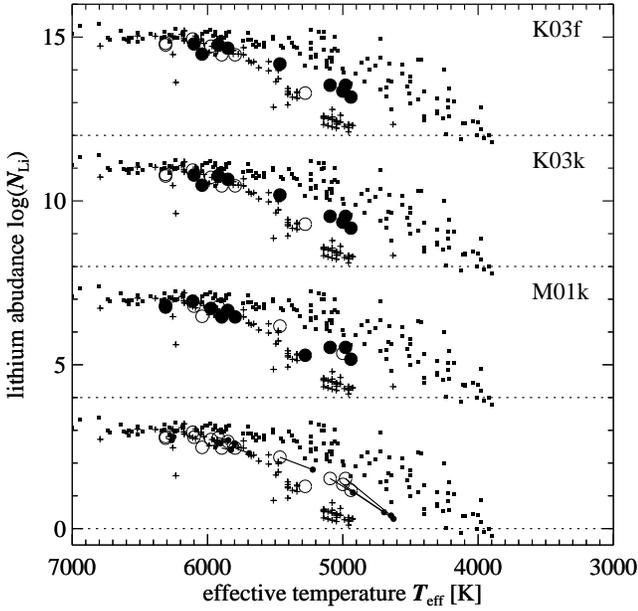}
\caption{\label{fig:teff_lognli}{\bf Effective temperature vs. lithium abundance} -- The values from Table~\ref{tab:memK_spec} (${\log}N^\mathrm{cog}_\mathrm{NLTE}(\mathrm{Li})$) are compared to data from \citet[][${\log}N^\mathrm{KS05}_\mathrm{LTE}(\mathrm{Li})$]{KS05} for stars in common, and {\bf to} the Hyades and Pleiades sequences which are taken from \citet{Sestito05b}. The layout of the bottom panel follows Fig.~\ref{fig:teff_ewli} a), and that of the upper three panels has the same meaning as Fig.~\ref{fig:teff_ewli} b). The upper three panels are shifted by arbitrary amounts with respect to the bottom panel. The dotted lines indicate an abundance level of zero to aid the eye.}
\end{figure}
\begin{figure}
\centering
\includegraphics[width=\columnwidth]{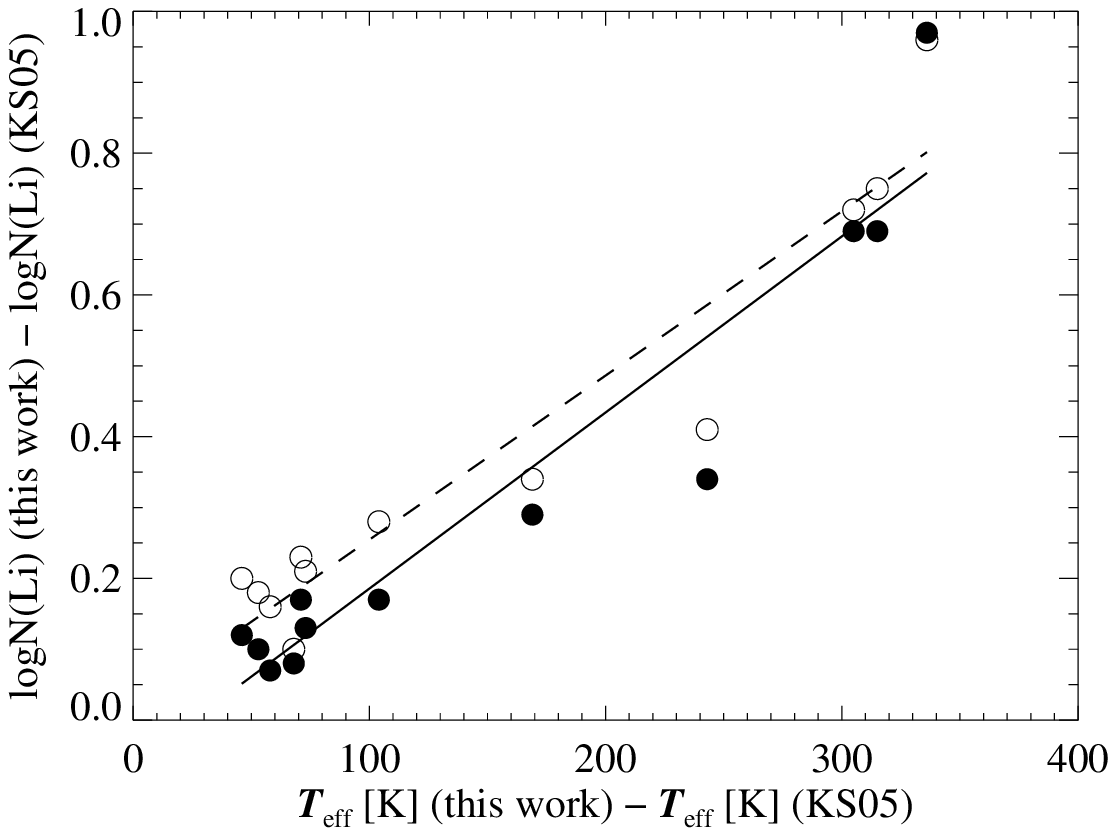}
\caption{\label{fig:dt_dnli}{\bf Systematics of measured Li abundances} -- for each star in common with \citet{KS05}, the difference of measured Li abundance is plotted versus the difference of adopted effective temperatures. The lines represent linear fits to the data. The values of \citet{KS05} given in Col.~9 of Table~\ref{tab:memK_spec} (${\log}N^\mathrm{KS05}_\mathrm{LTE}(\mathrm{Li})$) are subtracted from the LTE values presented in Col.~6 (${\log}N^\mathrm{cog}_\mathrm{LTE}(\mathrm{Li})$, filled circles, solid line) and Col.~8 (${\log}N^\mathrm{fit}_\mathrm{LTE}(\mathrm{Li})$, circles, dashed line). They are not compared to those of Col.~7, which were corrected for deviations from LTE and are used later. See also Fig.~\ref{fig:comp_Teff} for the systematic differences in effective temperature.}
\end{figure}

Table~\ref{tab:memK_spec} tabulates lithium abundances and equivalent widths measured for the Li\,I $\lambda$6708\,{\AA} feature. The observed spectroscopic data is supplemented by spectra kindly provided by K. Fuhrmann for HD\,11131, HD\,26923, HD\,39587, HD\,41593, and HD\,59747.

The measurements follow two different approaches. Firstly, the lithium equivalent widths are assessed by integrating the observed profiles. Using these data, abundance is derived in the same way as \citet{Sestito05b}, i.e., by interpolation of the LTE curves of growth of \citet[table~2]{Soderblom93a} and by applying the non-LTE (NLTE) corrections given by \citet{Carlsson94}\footnote{Instead of interpolating their Table~1, the data rows for metal-rich dwarfs are adopted, that is for a surface gravity of ${\log}g=4.5$ and an iron abundance of [Fe/H]$=0$.}. Secondly, the lithium abundances are determined by fitting the observed lithium feature with a synthetic profile to check the consistency with the curve-of-growth method. The synthetic Li profile is calculated in the same way as the Fe and Mg lines described before using the line data in Table~\ref{tab:mg_li_linedata}. 
Equivalent widths are not needed within this second approach but the equivalent width of the matching synthetic profile is given to compare to the observed values. 
Lithium abundance is only provided for the stars with uniformly derived stellar parameters (Table~\ref{tab:uma_mem_par}), so that additional systematics are not introduced and the relative accuracy of the results among the sample is preserved.

The quality of the measured equivalent widths strongly depends on the resolution and the signal-to-noise ratio of the underlying spectra. In the present work, a signal-to-noise ratio of at least $\approx200$ with a resolving power of $\approx60\,000$ was achievable for almost all stars. The blend with the Fe\,I line is mostly resolved. Therefore, the present equivalent widths of the UMa group can be considered to be very accurate. Uncertainties in the equivalent widths from direct integration are found by varying the integration limits and the position of the continuum, which depends on both noise and blends with neighbouring lines. Uncertainties in the curve-of-growth abundances are assessed by varying the input parameters, i.e., the equivalent widths, effective temperatures, and surface gravities within their error bars. The resulting total formal error amounts to typically less than $\approx0.10\,$dex. 

Similarly, the synthetic line fits are repeated by varying the stellar parameters within their error bars. The variation in both effective temperature and iron abundance dominated the changes in lithium abundance, so that only variations in these two parameters are considered. The final error again generally equals to $\approx0.10\,$dex.

The equivalent widths emerging from these methods agree remarkably well, except in the case of HD\,115043. The abundances from the LTE curves-of-growth should be consistent with the values obtained by the LTE line synthesis. Considering the problem of line blends, \citet{Huensch04} also derived Li abundance with both methods of synthetic line fits and curves of growth. Following their Fig.~3, the values from synthetic fits exceeded those from curves of growth \citep{Soderblom93a} by $0.1-0.3\,$dex. This effect can be reproduced from Table~\ref{tab:memK_spec}, although it is lower in magnitude. 
In the present work, the average difference $\left<{\log}N^\mathrm{cog}_\mathrm{LTE}(\mathrm{Li})-{\log}N^\mathrm{fit}_\mathrm{LTE} (\mathrm{Li})\right>=-0.05\pm0.03$ can be fully explained by the slightly sub-solar mean iron abundance of the UMa group and the neglected influence of iron abundance in the curve of growth tables of \citet{Soderblom93a}. Completing line synthesis with increased iron abundance shows that Li abundance needs to be reduced by approximately the same amount to fit the observed profile. 
Yet, a comparison with NLTE curve-of-growth results indicates that further systematics may be at work. \citet{Sestito05b} also derived Li abundance using the curves of growth of \citet{Soderblom93a} and the non-LTE corrections of \citet{Carlsson94}, and noted a similar effect of the same order when comparing with measurements for the Hyades (their Table~2). However, considering the previous explanations, an opposite effect would be expected since the Hyades have a super-solar iron abundance. 


The strength of the Li absorption and size of the Li abundance are correlated with effective temperature (Figs.~\ref{fig:teff_ewli} and \ref{fig:teff_lognli}). Effective temperatures are obtained from Table~\ref{tab:uma_mem_par} if available and otherwise taken from the literature as indicated in Table~\ref{tab:memK_spec}. The spectral type of HD\,155674\,A is indicative of an effective temperature of 5250\,K according to \citet{KH95}, whereas the H$\alpha$ wings favour a temperature as cool as 4800\,K or even lower.


The UMa group data is compared to the comprehensive compilation of \citet{KS05}, who obtained Li equivalent widths and abundances of UMa group members from both the literature and their own measurements (Table~\ref{tab:memK_spec}, Cols.~5 and 9, Figs.~\ref{fig:teff_ewli} and \ref{fig:teff_lognli}). The integrated equivalent widths $W^\mathrm{int}_\mathrm{Li}$ generally agree well with those given by \citet{KS05} ($W^\mathrm{KS05}_\mathrm{Li}$). The average difference amounts to $1.6\pm7.4\,\mathrm{m\AA}$, which is well within the error bars. Only in the case of HD\,109647, the values differ by as much as $21\,\mathrm{m\AA}$. \citet{KS05} adopted a Li equivalent width of $25\,\mathrm{m\AA}$ for this star from \citet{Soderblom93c}. The inspection of Table~1 of the latter shows that it is the only star with substantially conflicting measurements. The given values range from 14 to $40\,\mathrm{m\AA}$, the latter being in good agreement with the present work. 

The values of the present work are not only compared to previous measurements of the UMa group. Furthermore, they are studied relative to other stellar groups in a way as consistently as possible. Therefore, the UMa group is compared to the Pleiades and Hyades using data provided by \citet{Sestito05b} (Figs.~\ref{fig:teff_ewli} and \ref{fig:teff_lognli}). Some important points have to be kept in mind.
%
%
%
%
%
Uncertainties in the effective temperature may shift the Li sequence considerably. Except for the coolest stars, the effective temperatures were derived in a very homogeneous way making the UMa sequence very reliable in a relative sense. Figures \ref{fig:teff_ewli} and \ref{fig:teff_lognli} reproduce the systematic offset from the effective temperature scale of \citet{KS05}, which increases towards cooler stars. One cannot exclude similar offsets from the temperature scale of \citet{Sestito05b}, who derived effective temperature from $B-V$ colour in a way similar to \citet{KS05}. 

The derived Li abundance is affected substantially by those temperature offsets. While the equivalent widths presented by \citet{KS05} and this work agree well (Fig.~\ref{fig:teff_ewli}), the temperature offsets imply an offset in Li abundance of corresponding strength. \citet{KS05} derived Li abundances by interpolating in curve-of-growth tables generated by the LTE analysis package MOOG. These values can be compared with the LTE values derived in the present work (Table~\ref{tab:memK_spec}, Cols.~6, 8). Figure~\ref{fig:dt_dnli} shows that differences strongly correlate with the systematics in effective temperature.
The effect was modelled qualitatively using the line synthesis tools of the present study. A decrease in effective temperature implies a decrease in Li abundance when fitting the same spectrum. Both the Li abundance and the temperature offsets thus conspire in a peculiar way (Fig.~\ref{fig:teff_lognli}) so that in the end, the position of the UMa Li sequence remains virtually unchanged with respect to the sequences of the Hyades and Pleiades.

Differences in metallicity are also known to strongly affect the derivation of Li abundance. While the present work accounts for iron abundance within the synthetic line formation, the Li abundances presented by \citet{Sestito05b} are not corrected for metallicity. This particularly concerns the more metal-rich Hyades \citep[cf.][]{Soderblom93c,KS05}. The effect of the overall enhanced opacity on the derivation of Li abundance by LTE line formation was estimated by rescaling the Li abundance of the UMa group member HD\,11131, i.e., changing [Fe/H] by $+0.23\,$dex to match the average abundance of Hyades members. To reproduce the Li line of this star, Li abundance had to be changed by $-0.23\,$ dex. Despite this significant variation, the position of HD\,11131 in Fig.~\ref{fig:teff_lognli} is hardly changed with respect to the Hyades sequence.

The distribution of the warmer UMa group members in Figs.~\ref{fig:teff_ewli} and \ref{fig:teff_lognli} overlaps with the distributions of the Pleiades and the Hyades. \citet{Soderblom93c} argue that the scatter of equivalent widths in the UMa group in this temperature range is real. They compare pairs of UMa group members with very similar effective temperatures and conclude that differences in Li abundance cannot be explained by uncertain effective temperatures. They also note the reduced scatter of equivalent widths in the cooler K dwarf regime around $5000\,$K compared to the Pleiades. This is attributed to the small numbers and supported by the argument that a random selection of the more numerous Pleiades within one temperature bin would also produce a very narrow range of equivalent widths. The increased scatter below 5200\,K noticed by \citet{KS05} is not observed in the present work, which is caused by the small numbers and the way in which the sample was selected. 

The cooler UMa group stars clearly have different properties from those in the Hyades and Pleiades, although the plot region around $5500\,$K is sparsely populated. In the cool regime, lithium absorption of young stars strongly correlates with the age \citep[e.g.,][for a comprehensive study]{Sestito05b}. The Pleiades have an age of roughly $110\,$Myrs \citep{Terndrup00}, while the Hyades are much older with an age of $\approx625\,$Myrs \citep{Perryman98}.
In the case of two stars, HD\,152863\,A and HD\,205435, only upper limits could be measured, which is probably related to their giant status. Although trends in equivalent widths do not necessarily reflect trends in Li abundance in an unambiguous way, the distribution of the cool UMa group dwarf members can be well distinguished from the Pleiades and the Hyades. This conclusion is generally true for Li abundance as well, although it is then based on a smaller number of stars. The comparison with the Hyades in the present work and also in \citet{KS05} benefits from the increased number of Hyades members with known Li abundance compared to \citet{Soderblom93c}.

The UMa group sequence turns out extremely sharp between $5000\,$K and solar temperatures when considering the kinematic ({\it K03k}) and final members ({\it K03f}) of \citetalias{King03}. The deviant HD\,41593 is part of the kinematic member list of \citetalias{Montes01b}. \citet{Soderblom93c} already noted that the Li abundance of this particular star is unusually low compared to the other UMa group members.




These conclusions are based on relatively few stars. However, they gain confidence because only stars with a high membership probability are considered. All unsure candidates are excluded from the present work. Thus, lithium equivalent width and abundance seem a valuable membership criterion for UMa group candidates cooler than the Sun. Down to an effective temperature of $\approx4900\,$K, the strength of Li absorption as well as Li abundance should be in-between values for the sequences of the Pleiades and the Hyades. In this temperature range, neither cool giants nor cooler dwarf members should have any measurable Li line.


\subsubsection{Infilling of the H$\alpha$ core}

\begin{figure}
\centering
\includegraphics[width=\columnwidth]{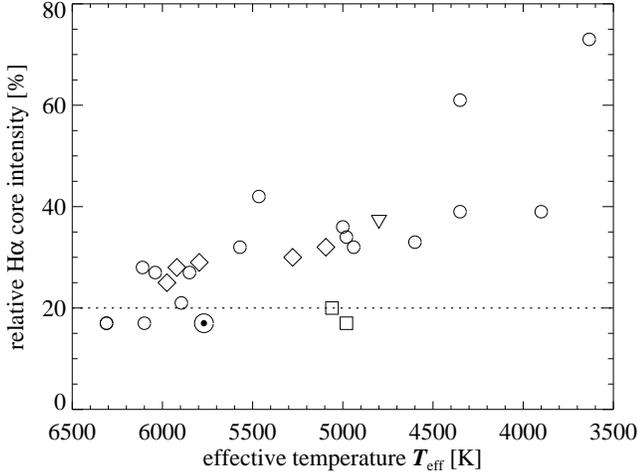}
\caption{\label{fig:teff_ha}{\bf Effective temperature vs. relative intensity of the H$\alpha$ core} -- 
The UMa group members (Table~\ref{tab:memK_spec}) are compared to the Sun ($\odot$). Diamonds indicate data taken from \citet{Fuhrmann04}, all other symbols  depict values derived in the present work. The squares denote the two giants HD\,152863\,A and HD 205435. HD\,155674\,A, which has an uncertain effective temperature, is indicated by the triangle. All Sun-like or cooler dwarf members of the UMa group have a relative intensity of the core of more than $20\,$\% (dashed line).}
\end{figure}

\begin{figure}
\centering
\includegraphics[width=\columnwidth]{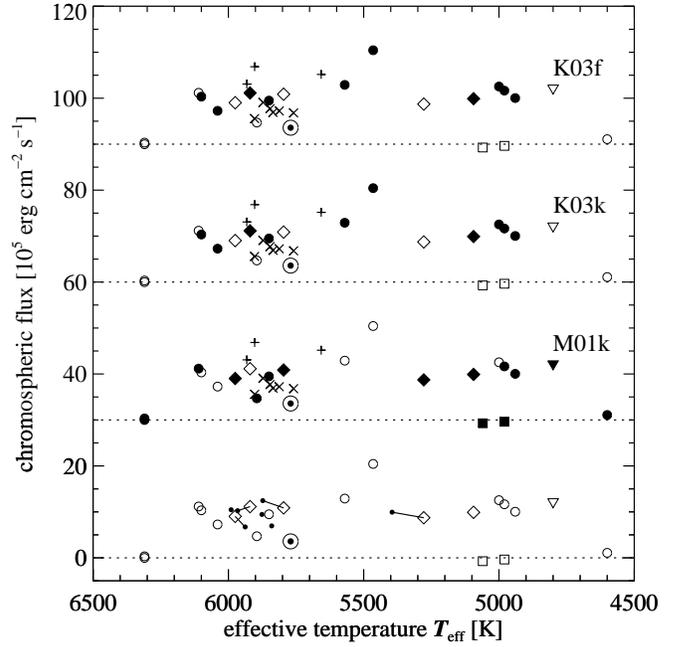}
\caption{\label{fig:plot_lyra}{\bf Effective temperature vs. chromospheric flux $F_\mathrm{H\alpha}$ \citep{Lyra05}} -- The layout corresponds to Fig.~\ref{fig:teff_lognli} indicating the three different samples {\it M01k}, {\it K03k}, and {\it K03f} by filled symbols in the upper three panels, which are shifted by arbitrary amounts with respect to the bottom panel. The dotted lines indicate a flux level of zero in order to aid the eye. Circles, squares, and diamonds indicate the measurements obtained in the present work. As in Fig.~\ref{fig:teff_ha}, the stars analysed by \citetalias{Fuhrmann04} are depicted by diamonds, the two giants HD\,152863\,A and HD\,205435 by squares, and HD\,155674\,A by the triangle. The measurements are compared to the Sun ($\odot$) and the measurements of \citet{Lyra05} for the Pleiades (plus signs), UMa group (dots), and the Hyades (crosses). The solid lines in the lowest panel connect measurements of the present work and \citet{Lyra05} for stars in common. There, the other data from \citet{Lyra05} are omitted for clarity.}
\end{figure}

\begin{figure}
\centering
\includegraphics[width=\columnwidth]{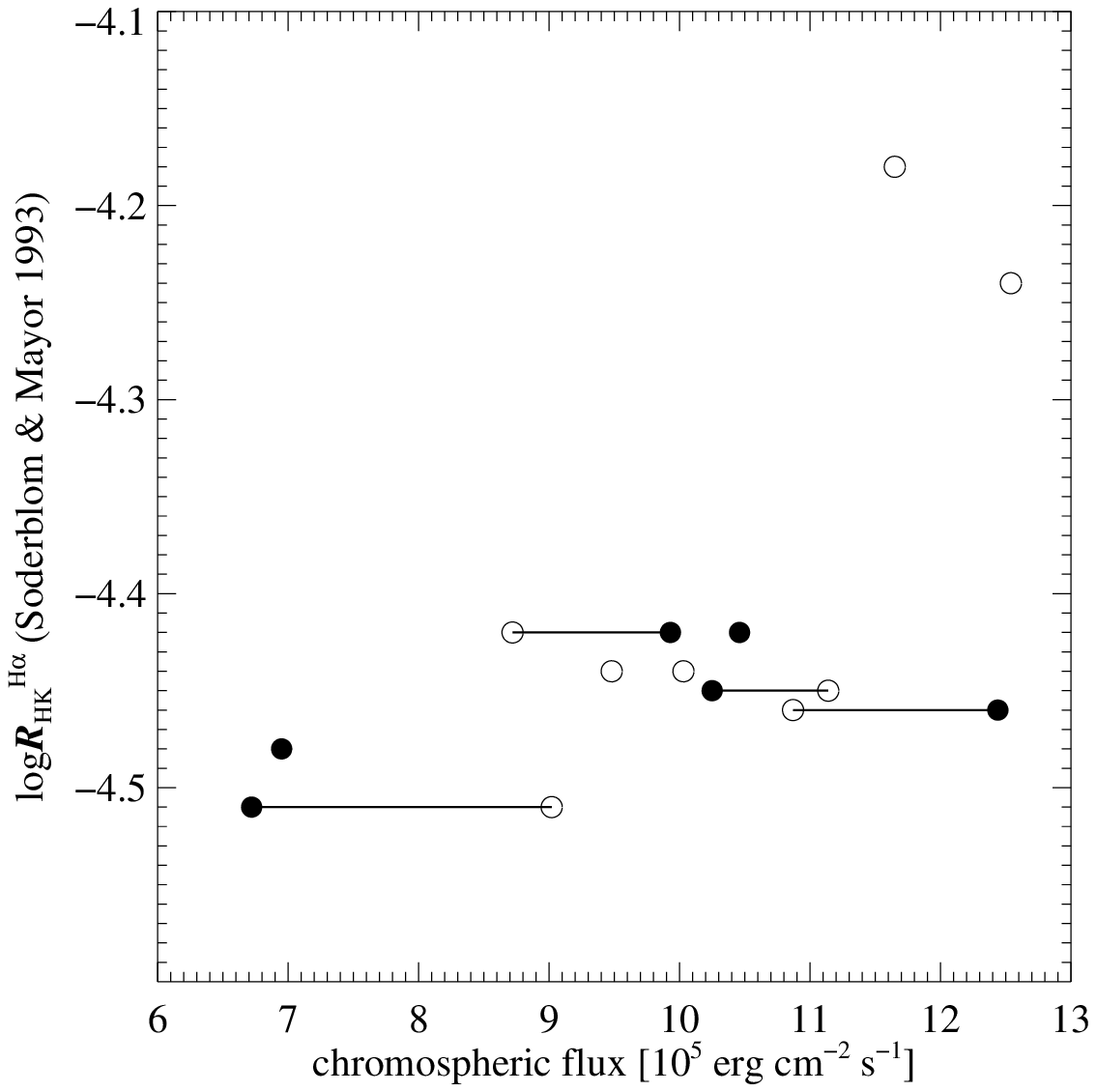}
\caption{\label{fig:plot_lyra_so} {\bf Chromospheric flux $F_\mathrm{H\alpha}$ vs. $R_\mathrm{HK}^\mathrm{H\alpha}$} -- The chromospheric emission index for H$\alpha$ of \citet{SM93} is compared to the chromospheric flux measure according to \citet{Lyra05} for UMa group stars in common. Empty circles indicate the chromospheric flux measurements of the present work, filled circles those of \citet{Lyra05}. Similar to Fig.~\ref{fig:plot_lyra}, the lines connect data points with different chromospheric flux measurements for the same star.}
\end{figure}

Chromospheric activity is usually quantified using indices based on Ca\,II\,H\&K and the Ca\,II infrared triplet (\citealp[see][for the UMa group]{SM93,King03}). Unfortunately, these lines are not present in the spectra available to the present work. However, it is possible to address the chromospheric infilling of the H$\alpha$ core. \citet{Lyra05} argue that there are even some advantages compared to Ca\,II H\&K.

\citet{Fuhrmann04} note that the H$\alpha$ cores of Ursa Major stars are filled-in by about $10\,$\% compared to inactive main-sequence stars of similar effective temperature. The present work allows one to verify this behaviour for a larger sample of UMa group members.

To be consistent with the work of \citet{Fuhrmann04}, the intensity level of the core of the H$\alpha$ line is assessed (Table~\ref{tab:memK_spec}) and compared to the solar case. The values are simply obtained by taking the minimum flux between 6560 and 6565\,{\AA}.
In the case of HD\,11131, HD\,26923, HD\,39587, HD\,41593, and HD\,59747, the values given by \citet[][Fig.~16]{Fuhrmann04} are adopted.

Figure~\ref{fig:teff_ha} generally confirms the findings of \citet{Fuhrmann04}. The UMa group dwarf members display an H$\alpha$ core elevated by on average about $10\,$\%. A tight sequence extends towards the coolest temperatures and clearly shows an offset with respect to the inactive solar case. The cool deviating objects are the two giants and a few outliers with obviously enhanced chromospheric activity (HD\,28495, HD\,238224, and HD\,95650). The samples {\it M01k}, {\it K03k}, and {\it K03f} (according to Table~\ref{tab:uma_mem_data}) tend to follow the same behaviour. As for the Li distribution, the identification of any deviating trends is limited by the small numbers of data points.

Figure~\ref{fig:teff_ha} allows one to conclude that no Sun-like or cooler UMa dwarf member is expected with a relative H$\alpha$ core intensity of less than about $20\,$\%. The definition of an upper limit, however, is found to be far less straightforward since individual members may have a significantly enhanced infilling.


To more reliably compare the chromospheric infilling of UMa group members with that of other stellar groups, in particular the Pleiades and the Hyades, the chromospheric flux is measured according to the method of \citet{Lyra05}. They showed that members of different stellar groups display a different average level of chromospheric flux.

Our measurements, which are produced with the method of \citet{Lyra05}, are shown in Table~\ref{tab:memK_spec}. The normalized spectra are integrated in both the H$\alpha$ core ($W_\mathrm{H\alpha}$) with a width of $1.7\,{\AA}$ and Willstrop's band ($6550-6600\,{\AA}$; $W_{50}$). The absolute flux in the H$\alpha$ core $F_\mathrm{H\alpha}$ is calculated from the ratio of $W_\mathrm{H\alpha}$ to $W_{50}$, using the absolute flux calibration and the $(V-R)-T_\mathrm{eff}$ calibration given by \citet{Lyra05}. Finally, photospheric flux is subtracted from $F_\mathrm{H\alpha}$ using the lower envelope in their Fig.~3\footnote{The coefficients of the spline were kindly provided by W. Lyra.}. Negative values close to zero are due to the intrinsic uncertainty on the photospheric flux corrections since the true photosperic flux is not known \citep[for a discussion see][]{Lyra05}. The calibrations provided by \citet{Lyra05} are only valid at effective temperatures as low as $5000\,$K. An extrapolation to the coolest stars in the present sample yields unrealistic results, which are excluded here (HD\,95650, HD\,155674\,B, HD\,238087, HD 238224). Figure~\ref{fig:plot_lyra} compares the values obtained in the present work to the values of Pleiades, UMa group, and Hyades members assessed by \citet{Lyra05}. The chromospheric flux of the Sun is measured from the exposures of the lunar disk obtained for the present work. On average, it is $3.58\pm0.35$, in excellent agreement with the value given by \citet{Lyra05} ($3.44\pm0.34$). At solar temperatures, the average flux levels of Pleiades, UMa group, and Hyades can be distinguished well, the Pleiades showing on average the largest and the Hyades having the lowest values. 

The bottom panel of Fig.~\ref{fig:plot_lyra} compares the measurements of stars common to both studies. For HD\,11131, HD 26923, HD\,39587, and HD\,41593, \citet{Lyra05} give the fluxes 12.44, 6.72, 10.25, and 9.93\,[$10^5$\,erg\,cm$^{-2}$\,s$^{-1}$]. The average difference from the measurements in the present work is $0.88\pm1.51$\,[$10^5$\,erg\,cm$^{-2}$\,s$^{-1}$]. On the one hand, this discrepancy is small compared to the total range of possible values covering three orders of magnitude. On the other hand, Pleiades, UMa group, and Hyades display fluxes of the same order of magnitude, so that the slight discrepancies between the present work and that of \citet{Lyra05} may become crucial.

\citet{SM93} assessed a chromospheric emission index of H$\alpha$ by subtracting the profile of an inactive star and obtaining the equivalent widths. These were converted to chromospheric flux ratios and rescaled to be compared to the index $R'_\mathrm{HK}$, which is based on the Ca\,II H\&K lines. These values ($R^\mathrm{H\alpha}_\mathrm{HK}$) are displayed in Fig.~\ref{fig:plot_lyra_so} for the stars in common with both the present work and \citet{Lyra05}. The work of \citet{SM93} has eight stars in common with the present work and six with \citet{Lyra05}, but only four stars are common to all three studies (indicated by solid lines in Fig.~\ref{fig:plot_lyra_so}). The comparison with \citet{Lyra05} allows for the inclusion of HD\,26913 and HD\,147513, which are not studied in the present work. 

Some stars in common could not be included. \citet{SM93} only determined an upper limit to the value of $R^\mathrm{H\alpha}_\mathrm{HK}$ for HD\,184960, and there is no such measurement at all for HD\,38393 and HD\,155674\,A and B. HD\,184960 was designated a probable spectrosopic non-member by \citet{SM93} while HD\,38393 and HD\,155674\,A and B are possible spectroscopic members. However, \citetalias{King03} classify HD\,38393 and HD\,184960 as activity-based non-members, while no activity indicator at all was assessed for HD\,155674\,A and B, inferring that they are additional UMa candiates. The present work assigns HD\,155674\,A an UMa-like chromospheric emission measure of 11.87, whereas HD\,155674\,B was excluded from the analysis. HD\,38393 and HD\,184960 do not exhibit any chromospheric H$\alpha$ flux. These two objects are the two hottest stars studied here, and low chromospheric emission is possibly common among these warm UMa group members. At these effective temperatures, \citet{Lyra05} did not study any Hyades stars, so that a more detailed comparison with \citet{SM93} and \citetalias{King03} cannot be made here.

It can be noticed from Fig.~\ref{fig:plot_lyra_so} that the different chromospheric H$\alpha$ emission measures behave in an expected way. They are correlated although substantial scatter is present. All objects in common that showed UMa-like chromospheric H$\alpha$ emission in \citet{SM93}, i.e., chromospheric indices larger than the Hyades mean relation, also have a chromospheric flux measure that is larger than 6.0 in the present work as well as in \citet{Lyra05}. This flux level separates the young and the old dwarf stars in \citet[][their Fig.~10]{Lyra05}.
\citet[][fig.~2]{SM93b} compared the H$\alpha$ emission of UMa group members to the Pleiades. The hotter UMa group stars overlap with the lower boundary of the wide distribution of the Pleiades. This overlap is also generally found in the present work and the work of \citet{Lyra05}, although represented by only three Pleiades stars.

In contrast to \citet{Lyra05}, the present work allows one to study the dependence of the chromospheric flux for a large range of effective temperature, i.e., stellar mass. Noticable flux levels are only measured at temperatures of $6000\,$K and below. At solar temperatures, a dependence on age is clearly noticed \citep{Lyra05}. The two UMa group giants HD\,152863\,A and HD\,205435 exhibit a significantly lower chromospheric flux level compared to dwarf stars of similar temperature.

In the temperature range from 6000\,K down to 5000\,K, the difference between Pleiades, UMa group, and Hyades is no longer so clear as in the case of only the solar temperature dwarfs.  The {\it K03f} sample displays chromospheric emission at the level of the Hyades or larger, in agreement with the spectroscopic criteria applied by \citet{SM93} and \citetalias{King03}. If only the dwarf stars cooler than $\approx$\,6000\,K are considered, then the chromospheric emission of all three samples {\it M01k, K03k}, and {\it K03f} is similar to the Hyades or stronger. The only exception is HD\,24916\,A, the coolest star with chromospheric emission measured in the present work. The spectroscopic criterion should thus be restricted to stars hotter than 5000\,K (on the effective temperature scale of the present work).

\section{Summary and conclusions}

Stellar parameters of 11 Ursa Major (UMa) group members have been derived homogeneously from spectra of both high resolving power and high signal-to-noise ratio using the methods of \citet[and preceeding work]{Fuhrmann04}. The method was only partially successful in the case of four stars so that surface gravities had to be calculated from the {\em Hipparcos} distance. 

The consistency of the analysis was proven by reproducing data for the Sun and a possible UMa group member already analysed by \citet{Fuhrmann04}, and by comparing the spectroscopic distance with the geometric {\em Hipparcos} distance. Furthermore, the derived values were found to be in good agreement with literature values, although some systematics were found relative to \citet{KS05}, who also analysed a larger sample of UMa group members.

To study the sample of UMa group members as a whole, the data set was complemented with data for 6 more UMa group members analysed with the same methods by \citet{Fuhrmann04} and \citet{Koenig06}. An iron abundance
of [Fe/H]$=-0.03\pm0.05$ was obtained for the UMa group, which is about $0.06\,$dex
higher than previous estimates. The [Fe/H] abundance is thus
indistinguishable from that of the Pleiades but clearly different from
that of the Hyades. The iron and magnesium abundances of the UMa group
members correspond to those of the thin disk population.

The equivalent widths of the Li\,I\,$6708\mathrm{\AA}$ feature were measured for in total 25 UMa group members and Li abundance for 16 of these stars. At effective temperatures lower than solar, the strength of the Li line and the Li abundance can be clearly distinguished from that of the Pleiades and the Hyades. 

Chromospheric emission was considered by two different approaches. The flux level of the H$\alpha$ core of 25 objects was measured according to \citet{Fuhrmann04}. In the case of the Sun-like and cooler UMa group members, it is elevated by about $10\,$\% with respect to the Sun, a result which is in line with previous findings by \citet{Fuhrmann04}. While the lower boundary of chromospheric infilling seems clearly defined, substantially higher values were found for some individual stars. 
Furthermore, chromospheric flux is calculated according to \citet{Lyra05} for 21 sample stars hotter than $5000\,$K. While the emission level is weaker than that of the Pleiades and stronger than that of the Hyades at solar temperatures, the distributions partly overlap at cooler temperatures. The chromospheric flux of all UMa group dwarfs between effective temperatures of 5000\,K and 6000\,K remains similar to that of the Hyades or is higher.

In previous works, it was found to be rather difficult to establish a definite list of bona-fide Ursa Major group members. Therefore, a simplistic approach was chosen in the present work by not trying to create a single and definite list that would be the basis of the definition of spectroscopic membership criteria. Different lists from the literature were adopted instead by considering the distribution of iron abundance, Li absorption and H$\alpha$ filling-in separately for each of these lists. A qualitative difference of the results of the samples cannot be assessed robustly since the number of stars studied is small. The kinematic and final members of \citet{King03} form remarkably tight sequences of Li equivalent width and abundance and a fairly homogeneous level of the chromospheric H$\alpha$ emission measure.

The purely kinematically selected stars appear similar in terms of their spectroscopic properties. In particular, the spread of iron abundance is much smaller than the total spread for the thin disk population. Most notably, the scatter of iron abundance is of the same order as the typical abundance measurement error bar. Similarly, the scatter of Li abundance is small along a temperature-dependent sequence and clearly differs from that of other groups of stars, supporting a young age for the UMa group in-between that for the Pleiades and the Hyades. Chromospheric emission in the H$\alpha$ line core is also indicative of youth since it is similar to the Hyades. Thus, the present study corroborates previous findings that the existence of the UMa group does not depend entirly on unsure kinematic criteria alone. The present work suggests that spectroscopic criteria may help to identify additional UMa group members.

Membership criteria derived from the mean iron abundance, the Li absorption, and the H$\alpha$ infilling can be used to exclude non-members \citep{King03}, i.e., the fulfillment of these criteria is necessary but insufficient for deciding on membership. UMa group members should have an iron abundance [Fe/H] in the range $[-0.14,0.06]$. Members cooler than the Sun and hotter than $5300$\,K are expected to have stronger Li absorption than the Hyades but weaker than $100\,\mathrm{m{\AA}}$. The Li feature should be present in the range $5300-4900\,$K but weaker than $50\,\mathrm{m{\AA}}$. Giants and cooler dwarfs in the UMa group are not expected to have any measurable Li\,I $6708${\AA} feature. The flux level in the H$\alpha$ core at effective temperatures between 5000\,K and 6000\,K is similar to that of the Hyades or higher. There is no measurable infilling for the hotter stars and the giants. Chromospheric emission may reach extreme levels in the case of the coolest UMa group members.

These criteria may help us to exclude non-members but cannot be used as a membership criterion in a positive sense. For this purpose, a complete census of stellar kinematic groups in the solar neighbourhood has to be assessed. To enable an accurate comparison, possible systematics have to be removed, which may in turn originate in inhomogeneous data sets. 

The presented spectroscopic criteria may certainly help us to identify possible members of the UMa group among the kinematic candidates. For this purpose, new candidates have to be analysed spectroscopically in the same way as in the present work or at least in a very similar way. Then, chemical abundance, in particular Fe and Li, as well as chromospheric infilling can be addressed by the spectroscopic criteria in a very uniform way.

The reader may have noticed that the spectroscopic criteria are based on only very few kinematic members, and will have to be placed on firmer grounds. Therefore, the first challenge will be to follow \citet{SM93}, \citet{King03}, and \citet{Fuhrmann04} in studying whether new spectroscopic UMa group candidates also cluster tightly in kinematic space. In this way, the kinematic membership criteria can be refined and more stars included in the definition of spectroscopic criteria. A larger list of members can then be established in a more reliable way.

\begin{acknowledgements}
 M.A. wants to thank Ralph Neuh\"auser and
Gregor. E. Morfill who supervised and supported the PhD thesis. M.A. is indebted to Klaus Fuhrmann who kindly provided some of his FOCES spectra and his analysis package and gives ongoing support. Furthermore, M.A. likes to thank Brigitte K\"onig for help in the analysis of young stars and Andreas Korn for software to add spectra. Moreover, M.A. gratefully acknowledges fruitful discussions with Giancarlo Pace, G\"unter Wuchterl, and Marc Hempel. M.A. and E.G. thank the support staff at the observatories at Calar Alto and at Tautenburg. We thank the anonymous referee for comments and suggestions which clearly helped to improve this work.
M.A. was supported by a graduate scholarship of the Cusanuswerk, one of the national student elite programs of
Germany and by a scholarship (reference SFRH/BPD/26817/2006) granted by the {\it Funda\c{c}\~ao para a Ci\^{e}ncia e a Tecnologia} (FCT), Portugal. M.A. acknowledges research funding granted by the {\it Deutsche Forschungsgemeinschaft} (DFG) under the
projects NE 515/16 and 17, and RE 1664/4-1. This research has made use of the {\it extract stellar} request type of the Vienna Atomic Line Database (VALD), the SIMBAD database, operated at CDS, Strasbourg, France, and NASA's Astrophysics Data System Bibliographic Services.
\end{acknowledgements}

\bibliographystyle{aa}
\bibliography{./uma_par}

\Online

\begin{appendix}
\section{List of Fe lines used in this work}
\begin{table*}
\caption[List of Fe lines used in this work]{\label{tab:qline_fe_used}Data of Fe\,I and II lines used in this work.}
\begin{tabular}{cccccl|cccccl}
\toprule
$\lambda$ [\AA]&ion. stage&$\chi_\mathrm{low}$ [eV]&${\log}gf$&${\log}C_6$&$\gamma_\mathrm{rad}$ [$10^8$\,1/s]&$\lambda$ [\AA]&ion. stage&$\chi_\mathrm{low} [eV]$&${\log}gf$&${\log}C_6$&$\gamma_\mathrm{rad}$ [$10^8$\,1/s]\\
\midrule
     4993.35&  2& 2.80& -3.73& -31.60& 3.0500&      6252.57&  1& 2.40& -1.56& -31.92& 0.87000\\     
     4994.14&  1& 0.91& -3.01& -31.97& 0.15000&     6297.80&  1& 2.22& -2.71& -31.86& 1.5500\\      
     5127.37&  1& 0.91& -3.20& -31.97& 0.16600&     6301.51&  1& 3.65& -0.67& -30.70& 0.80500\\     
     5197.58&  2& 3.23& -2.31& -31.80& 3.0300&      6330.85&  1& 4.73& -1.15& -31.69& 2.3840\\      
     5198.72&  1& 2.22& -2.12& -31.50& 1.6100&      6335.34&  1& 2.20& -2.22& -31.86& 1.6600\\      
     5217.40&  1& 3.21& -1.08& -30.60& 0.76000&     6336.83&  1& 3.69& -0.80& -30.70& 0.81100\\     
     5223.19&  1& 3.64& -2.25& -31.91& 0.78600&     6369.46&  2& 2.89& -4.19& -31.50& 2.9000\\      
     5234.63&  2& 3.22& -2.23& -31.80& 3.0700&      6380.75&  1& 4.19& -1.24& -31.70& 0.73503\\     
     5242.50&  1& 3.63& -0.87& -31.70& 0.57500&     6393.61&  1& 2.43& -1.39& -31.92& 0.93200\\     
     5264.81&  2& 3.23& -3.11& -32.19& 4.1100&      6411.66&  1& 3.65& -0.56& -30.70& 0.80300\\     
     5281.80&  1& 3.04& -0.91& -30.90& 1.0200&      6416.93&  2& 3.89& -2.67& -31.90& 3.3700\\      
     5284.11&  2& 2.89& -3.12& -32.11& 3.4300&      6430.85&  1& 2.18& -2.02& -31.86& 1.6500\\      
     5295.32&  1& 4.42& -1.52& -31.15& 1.8380&      6432.68&  2& 2.89& -3.63& -31.50& 2.9000\\      
     5302.31&  1& 3.28& -0.75& -30.60& 0.74900&     6436.41&  1& 4.19& -2.39& -31.96& 0.30403\\     
     5325.56&  2& 3.22& -3.22& -31.80& 3.1000&      6456.39&  2& 3.90& -2.09& -31.90& 3.3700\\      
     5339.94&  1& 3.27& -0.65& -30.60& 0.74900&     6481.88&  1& 2.28& -2.90& -31.91& 1.5500\\      
     5364.88&  1& 4.45&  0.39& -31.15& 1.8550&      6498.94&  1& 0.96& -4.62& -32.02& 0.000435\\    
     5367.48&  1& 4.42&  0.49& -31.15& 1.8500&      6516.09&  2& 2.89& -3.34& -31.50& 2.9100\\      
     5379.58&  1& 3.69& -1.44& -31.70& 0.57500&     6591.32&  1& 4.59& -2.01& -31.80& 1.4000\\      
     5393.17&  1& 3.24& -0.74& -30.60& 0.74800&     6593.88&  1& 2.43& -2.29& -31.92& 0.86200\\     
     5414.07&  2& 3.22& -3.61& -32.19& 4.1200&      6608.03&  1& 2.28& -3.99& -31.91& 1.6600\\      
     5425.25&  2& 3.20& -3.28& -31.80& 2.9900&      6627.55&  1& 4.55& -1.48& -31.44& 1.7530\\      
     5497.52&  1& 1.01& -2.82& -31.99& 0.14200&     6704.48&  1& 4.22& -2.62& -31.16& --\\          
     5522.45&  1& 4.21& -1.40& -30.80& 0.89800&     6710.32&  1& 1.48& -4.84& -31.97& 0.16600\\     
     5534.85&  2& 3.24& -2.83& -32.18& 2.9900&      6713.75&  1& 4.80& -1.42& -31.30& 2.3550\\      
     5546.51&  1& 4.37& -1.05& -31.20& 1.7860&      6725.36&  1& 4.10& -2.21& -31.37& 2.0910\\      
     5560.22&  1& 4.43& -0.96& -31.50& 1.6410&      6733.15&  1& 4.64& -1.43& -31.37& 2.2680\\      
     5576.10&  1& 3.43& -0.79& -30.85& 0.75300&     6750.16&  1& 2.42& -2.53& -31.95& 0.0770\\      
     5611.36&  1& 3.64& -2.93& -31.91& 1.2500&      6786.87&  1& 4.19& -1.90& -31.36& 1.9860\\      
     5618.64&  1& 4.21& -1.22& -31.46& 1.0450&      6806.85&  1& 2.73& -3.11& -31.83& 1.0200\\      
     5636.70&  1& 3.64& -2.52& -31.94& 0.39000&     6810.27&  1& 4.61& -0.94& -30.90& 2.3010\\      
     5638.27&  1& 4.22& -0.65& -31.00& 1.9430&      6820.37&  1& 4.64& -1.11& -30.90& 2.2200\\      
     5651.47&  1& 4.47& -1.77& -31.30& 1.6310&      6837.01&  1& 4.59& -1.74& -31.81& 0.52405\\     
     5679.03&  1& 4.65& -0.64& -30.80& 1.3990&      6843.66&  1& 4.55& -0.73& -31.10& 1.9230\\      
     5741.85&  1& 4.26& -1.63& -31.10& 2.1080&      6857.25&  1& 4.08& -2.08& -31.40& 0.25303\\     
     5855.08&  1& 4.61& -1.54& -31.75& 1.9110&      6862.50&  1& 4.56& -1.43& -31.30& 3.5400\\      
     5856.09&  1& 4.29& -1.55& -31.91& 0.86200&     6936.50&  1& 4.61& -2.20& -31.69& 3.2000\\      
     5862.36&  1& 4.55& -0.16& -31.30& 1.8970&      6945.21&  1& 2.42& -2.39& -31.95& 0.15700\\     
     5930.19&  1& 4.65& -0.02& -31.30& 1.8530&      6978.86&  1& 2.48& -2.39& -31.95& 0.07700\\     
     5956.70&  1& 0.86& -4.54& -32.01& 0.000271&    7219.69&  1& 4.08& -1.49& -31.91& 1.6300\\      
     5991.38&  2& 3.15& -3.59& -31.60& 3.4600&      7222.40&  2& 3.89& -3.26& -31.60& 4.1400\\      
     6027.06&  1& 4.08& -1.02& -31.91& 0.88503&     7224.48&  2& 3.89& -3.24& -31.60& 4.1400\\      
     6056.01&  1& 4.73& -0.31& -30.80& 1.8480&      7301.57&  2& 3.89& -3.69& -32.18& 3.0700\\      
     6065.49&  1& 2.61& -1.38& -31.92& 1.0200&      7306.57&  1& 4.18& -1.54& -30.80& 1.6710\\      
     6084.11&  2& 3.20& -3.84& -31.60& 3.4300&      7401.69&  1& 4.19& -1.52& -31.90& 0.70303\\     
     6093.65&  1& 4.61& -1.32& -31.75& 1.9350&      7411.16&  1& 4.28& -0.32& -30.80& 1.7020\\      
     6096.67&  1& 3.98& -1.79& -31.10& 0.45250&     7421.56&  1& 4.64& -1.70& -31.82& 2.5010\\      
     6113.32&  2& 3.22& -4.16& -31.60& 3.4100&      7445.76&  1& 4.26& -0.10& -30.80& 1.6990\\      
     6149.25&  2& 3.89& -2.75& -31.90& 3.3900&      7454.00&  1& 4.19& -2.34& -31.90& 1.4500\\      
     6151.62&  1& 2.18& -3.29& -31.86& 1.5500&      7479.70&  2& 3.89& -3.70& -32.18& 3.1000\\      
     6157.73&  1& 4.08& -1.10& -31.70& 0.50203&     7491.66&  1& 4.30& -0.95& -30.80& 1.6920\\      
     6173.34&  1& 2.22& -2.84& -31.86& 1.6700&      7495.08&  1& 4.22&  0.04& -30.80& 1.6810\\      
     6200.32&  1& 2.61& -2.31& -31.92& 1.0300&      7498.53&  1& 4.14& -2.16& -31.90& 1.1600\\      
     6213.44&  1& 2.22& -2.51& -31.86& 1.6500&      7515.84&  2& 3.90& -3.45& -31.60& 4.0900\\      
     6226.74&  1& 3.88& -2.11& -31.10& 0.54140&     7540.44&  1& 2.72& -3.81& -31.85& 0.93200\\     
     6232.65&  1& 3.65& -1.20& -30.70& 0.80600&     7568.91&  1& 4.28& -0.82& -30.80& 1.6990\\      
     6239.94&  2& 3.89& -3.50& -31.90& 3.3800&      7583.80&  1& 3.02& -1.80& -31.80& 1.0200\\      
     6240.65&  1& 2.22& -3.27& -31.95& 0.15700&     7711.73&  2& 3.90& -2.56& -31.60& 4.1200\\      
     6246.33&  1& 3.60& -0.73& -30.70& 0.79600&     7748.28&  1& 2.95& -1.56& -31.80& 1.0200\\      
     6247.56&  2& 3.89& -2.34& -31.90& 3.3800&     7844.56&  1& 4.83& -1.73& -31.59& 2.3380\\  
 
\bottomrule
\end{tabular}
\newline
{\bf (1)} central wavelength, {\bf (2)} ionization stage, {\bf (3)} lower excitation potential, {\bf (4)} oscillator strength, {\bf (5)} van der Waals damping constant, {\bf (6)} radiative damping constant. The effect of quadratic Stark broadening (${\log}C_4$) on weak iron lines is negligible so that in this work, it is not taken into account for these lines. The radiation damping constant ($\gamma_\mathrm{rad}$) is calculated with the classical formula in case it is missing in the table.\end{table*}

\end{appendix}

\end{document}